\documentclass[aps,twocolumn,eqsecnum,unsortedaddress]{revtex4}
\usepackage{amsmath,graphicx,epsfig}

\def\const{\mathrm{const\,}}
\def\sn{\mathrm{sn\,}}

\allowdisplaybreaks

\def\openone{\leavevmode\hbox{\small1\kern-3.3pt\normalsize1}}

\begin{document}

\arraycolsep=2pt
\bibliographystyle{apsrev}

\title{Adiabatic $N$-soliton interactions of Bose-Einstein condensates
in external potentials}

\author{V. S. Gerdjikov$^{1}$, B.\ B.\ Baizakov$^2$,
M.\ Salerno$^2$ and N.\ A.\ Kostov$^3$ }

\email[V.Gerdjikov:]{gerjikov@inrne.bas.bg}
\email[B.Baizakov:]{baizakov@sa.infn.it}
\email[M.Salerno:]{salerno@sa.infn.it}
\email[N.Kostov:]{nakostov@ie.bas.bg}

\affiliation{$^1$Institute for Nuclear Research and
Nuclear Energy, Bulgarian Academy of Sciences, Blvd. Tzarigradsko
chaussee 72, 1784 Sofia, Bulgaria.}

\affiliation{$^2$Dipartimento di Fisica ``E.R.Caianiello"
and Consorzio Nazionale Interuniversitario per le Scienze Fisiche
della Materia (CNISM) Unit\'a di Salerno,\\
Universit\'a di Salerno, via S. Allende, I84081 Baronissi (SA),
Italy;}

\affiliation{$^3$Institute for Electronics, Bulgarian Academy of
Sciences,  Blvd. Tzarigradsko chaussee 72, 1784 Sofia, Bulgaria}

\date{\today}

\arraycolsep=2pt

\begin{abstract}

 Perturbed version of the complex Toda chain (CTC) has been
employed to describe adiabatic interactions within $N$-soliton
train of the nonlinear Schr\"odinger equation (NLS). Perturbations
induced by weak quadratic and periodic external potentials are
studied by both analytical and numerical means. It is found that
the perturbed CTC adequately models the $N $-soliton train
dynamics for both types of potentials. As an application of the
developed theory we consider the dynamics of a train of matter -
wave solitons confined to a parabolic trap and optical lattice, as
well as tilted periodic potentials. In the last case  we
demonstrate that there exist critical values of the strength of
the linear potential for which one or more localized states can be
extracted from a soliton train. An analytical expression for these
critical strengths for expulsion is also derived.
\end{abstract}


\maketitle

\section{Introduction}

One of the most remarkable phenomena occurring in nonlinear
systems  is the possibility to form localized states of soliton
type as a result of the interplay  between dispersion and
nonlinearity. These states display interesting properties in
response to external fields and their interactions have been the
subject of continuous interest since the creation of the soliton
theory. Besides the motivation from the viewpoint of fundamental
physics, studies on soliton interactions are very important for
applications such as  optical fiber communications systems, where
optical solitons are used as information bit carriers
\cite{hasegawa}. In this context the interaction between
neighboring solitons in a train limits the transmission capacity
of the communication system, so that soliton interaction becomes
very important for optimal information packing and transmission
rates design. Trains of solitons (fluxons) in interaction play an
important role also in long Josephson junctions where their
dynamics, induced by external magnetic fields, is used to
construct flux-flow oscillators, devices of interest for
applications in superconducting mm and sub-mm wave electronics
\cite{barone}. Other  rapidly developing fields  in which
interacting solitons play a crucial role are photonic crystals
\cite{kivshar} and Bose-Einstein condensates (BEC).

Recently, BEC solitons have attracted a great deal of interest
both from the theoretical and the experimental point of view (see
e.g. reviews \cite{abd}). In particular, self-trapped states
capable to propagate in space without distortion are of interest
for pulsed atomic soliton lasers, atomic nano-litography and  high
precision interferometry \cite{jpb} and the transport of BEC
solitons in the presence of external potentials, serving as
magnetic and optical traps and waveguides may become important in
future technologies.

The aim of the present paper is to study the adiabatic dynamics of
a train of $N$ interacting  solitons of the nonlinear
Schr\"odinger equation (NLS) in weak external potentials. As a
physical model we consider matter-wave solitons in quasi-one
dimensional BEC with attractive interactions between atoms, such
as $^7$Li, $^{85}$Rb or $^{137}$Cs. The results, however, are of
interest also for nonlinear optics and for photonic crystals. In
particular, we study  the N-soliton dynamics in  parabolic, linear
and periodic potentials, modelling the NLS train soliton
interaction in terms of a complex Toda chain (CTC) which is valid
in the adiabatic approximation. This model has been successfully
used in previous papers (see Refs.
\cite{GKUE,GUED,Arnold,GKDU,GU,G_Gal02})  and will be used here to
continue the analysis in \cite{Liss,GBS} on the perturbed CTC
(PCTC) as a model for N-soliton interaction in BEC with quadratic
and periodic potentials. Our results provide an additional
confirmation of the stabilization properties of the periodic
potentials observed in \cite{WB,UGL} in a different physical
setup. In particular, for the case of an $N $-soliton train
trapped in a weak parabolic trap  we find that the train performs
contracting and expanding oscillations if its center of mass
coincides with the minimum of the potential, while it oscillates
around the minimum of the potential as a whole if its center of
mass is shifted from the minimum. In the last case contracting and
expanding motions of the soliton train is superimposed to the
center of mass dynamics. As the strength of the parabolic trap
increases  we find from numerical simulations that the N-soliton
dynamics becomes more complicated with the the merging (splitting)
of individual solitons when the train is contracted (expanded)
during its oscillating motion in the trap. This behavior resemble
the phenomenon of "missing solitons" observed in the experiment
\cite{strecker}.

We have investigated the case of a tilted periodic potentials i.e.
a potential which is the superposition of periodic and linear
potentials. The effect of the linear potential, if it is strong
enough, is that it can overcome the stabilization effect of the
periodic potential. As a result we show that there exist critical
values of the strength of the linear potential for which one or
more localized states can be extracted from a soliton train
(array). We find that the critical strength for expulsion changes
with the number of the solitons in the train. To this regard we
have derives an analytical expression for the potential critical
strengths at which expulsion is achieved by means of an
Hamiltonian approach for the PCTC. From this comparison we find
that the analysis based on the PCTC provides a good description
for the expulsion phenomena.

We remark that since the critical value of the potential strength
is an indirect measure of the binding energy of an N-solitons
train, one could use BEC soliton arrays in accelerated optical
lattices to measure N-solitons matter waves binding energies. One
could indeed reproduce the effect of the linear potential by means
of  accelerated optical lattices and measuring the critical
acceleration at which the expulsion of one soliton from the array
occurs. We hope experiments in this direction will be performed
soon.

The paper is organized as follows. In section II we introduce the
mean field Gross-Pitaevskii equation (GPE) appropriate for quasi
one dimensional  BEC in external potentials and discuss the
N-soliton interaction in terms of the complex Toda chain. In
Section III we study the perturbed nonlinear Schr\"odinger
equation and the perturbed CTC equation in the adiabatic
approximation for different types of external potentials:
quadratic, linear and periodic. The analysis of the N-soliton
dynamics obtained from the perturbed CTC with the above potentials
is compared with direct numerical GPE simulations in section IV.
In section V we introduce a Hamiltonian formulation of the PCTC
and derive an analytical expression for the critical strengths of
the linear potential for which the phenomenon of soliton expulsion
occurs. Finally, in the last section the main results of the paper
are briefly summarized.

\section{Model equations and N-soliton interaction}
The dynamics of a condensate in the mean-field approximation at
zero temperature is governed by the 3D nonlinear Schr\"odinger
equation (in the BEC context called also as Gross-Pitaevskii
equation (GPE))
\begin{equation}
i\hbar \frac{\partial \Psi }{\partial t}=\left[ -\frac{\hbar ^{2}}{2m}\nabla
^{2}+V(x,y,z)+\frac{4\pi \hbar ^{2}a_{s}{\cal N}}{m}|\Psi |^{2}\right] \Psi ,
\label{3Dgpe}
\end{equation}
where $\Psi (\mathbf{r},t)$ is the macroscopic wave function of the
condensate normalized so that $\int |\Psi
(\mathbf{r})|^{2}d\mathbf{r}=1$, ${\cal N} $ is the total number of
atoms, $m$ is the atomic mass, $a_{s}$ is the $s$-wave scattering
length (below we shall be concerned with attractive BEC for which
$a_{s}<0$), and
\begin{equation}
V(x,y,z)=\frac{m}{2}[\omega _{x}^{2}x^{2}+\omega _{\bot
}^{2}(y^{2}+z^{2})]
\end{equation}
is the axially symmetric trapping potential which provides for the
tight confinement in the transverse plane $\left( y,z\right) $, as
compared to loose axial trapping, assuming $\omega _{x}^{2}/\omega
_{\bot }^{2}\ll 1$. The condensate trapped in such a potential
acquires highly elongated form.

 When the transverse confinement is strong enough, so that the
transverse oscillation quantum, $\hbar \omega _{\bot }$, is much
greater than the characteristic mean-field interaction energy,
$(4\pi\hbar^2a_s/m){\cal N}|\Psi |^{2}$, the dynamics is
effectively one-dimensional. In this case, the wave function may
be effectively factorized as $\Psi (x,y,z,t)=\psi (x,t)\phi
(y,z)$, where $\phi (y,z)=\exp [-(y^{2}+z^{2})/2a_{\bot
}^{2}]/\sqrt{\pi }a_{\bot }$ is the normalized ground state of the
2D harmonic oscillator in the transverse direction, with $a_{\bot
}=\sqrt{\hbar /m\omega _{\bot }}$ being the corresponding
transverse harmonic-oscillator length. Inserting the factorized
expression into the 3D GPE (\ref{3Dgpe}), and integrating it over
the transverse plane $(y,z)$, one derives the effective 1D
equation
\begin{equation}
i\hbar \frac{\partial \psi }{\partial t}=\left[ -\frac{\hbar
^{2}}{2m}\frac{\partial ^{2}}{\partial x^{2}}+\frac{m}{2}\omega
_{x}^{2}x^{2}+g_{1D}{\cal N}|\psi |^{2}\right] \psi ,
\label{1Dgpe}
\end{equation}
where we have neglected the zero-point energy of the
transverse motion, $\hbar \omega _{\bot }$, and defined a
coefficient of the 1D nonlinearity, $g_{1D}=4\pi \hbar
^{2}a_{s}m^{-1}\int |\phi (y,z)|^{4}dydz=2a_{s}\hbar \omega _{\bot}$.
It is convenient to use normalized units for time and space variables,
introducing transformations:
$t\rightarrow \omega _{\bot}t$, $x\rightarrow x/a_{\bot}$,
and the rescaled wave function
$u \rightarrow \sqrt{2{\cal N} |a_{s}|}\psi $
\begin{equation}
i\frac{\partial u}{\partial t}+\frac{1}{2}\frac{\partial
^{2}u }{\partial x^{2}}-V_2 x^{2}u +|u |^{2}u =0,
\label{PDE}
\end{equation}
where $V_2 = \omega_x^2/(2\omega_{\bot}^2)$, is a small parameter
characterising the strength of the external parabolic potential.
In the experiment of Rice university \cite{strecker} a matter-wave
soliton train (comprising $N \sim 10$ solitons) of Bose-condensed
$^7$Li atoms ($a_s=-0.21$ nm, $m=11.65 \cdot 10^{-27}$ kg) was
created. Radial confinement was strong, $\omega_{\bot} \sim$ 800
Hz, while the axial one had been as weak as $\omega_x \sim$ 3 Hz.
In the experiment \cite{khaykovich}, where a single soliton of
$^7$Li BEC was created, the trap aspect ratio was smaller:
$\omega_{\bot} \sim$ 710 Hz, $\omega_x \sim$ 50 Hz. Therefore,
$V_2 \sim 10^{-5} \div 10^{-3}$ is in the range of realistic
experimental conditions. Below we shall consider also other kinds
of weak potentials in the axial direction $x$, instead of (or
combined with) the parabolic trap in Eq. (\ref{PDE}), see
\cite{BCCD,carretero}. Assuming them as perturbations
$iR[u]=V(x)u(x,t) $, we move them to the right hand side of the
governing equation.

The $N $-soliton train interactions for the nonlinear
Schr\"odinger equation (NLS) and its perturbed versions
\begin{equation}\label{eq:nls}
i\frac{\partial u}{\partial t}+\frac{1}{2}\frac{\partial
^{2}u }{\partial x^{2}} + |u |^{2}u = i\epsilon R[u],
\end{equation}
started with the pioneering paper \cite{Karp}, by now has been extensively
studied (see \cite{GKUE,GUED,Arnold,GKDU,Liss} and references therein).
Several other nonlinear evolution equations (NLEE) were also studied,
among them the modified NLS equation \cite{KodHas,IG_BJ,Val&I,GDY,GU},
some higher NLS equations \cite{Liss}, the Ablowitz-Ladik system
\cite{Dokt} and others.

Below we concentrate on the perturbed NLS eq. (\ref{eq:nls}).
By $N $-soliton train we mean a solution of the (perturbed) NLS fixed up
by the initial condition:
\begin{equation}\label{eq:IC}
u(x,t=0) = \sum_{k=1}^{N} u_k^{\rm 1s}(x,t=0) ,\qquad
u_k^{\rm 1s}(x,t) = {2\nu _k e^{i\phi _k}\over \cosh z_k },
\end{equation}
\begin{equation}\label{eq:nls-1s}
z_k(x,t) = 2\nu _k (x-\xi _k (t)), \qquad \xi_k(t) = 2\mu _kt +
\xi_{k,0}
\end{equation}
\begin{equation}
\phi _k(x,t)= {\mu _k \over \nu _k} z_k + \delta _{k}(t),\qquad
\delta _k(t) = W_kt +\delta _{k,0},
\end{equation}
Each soliton has four parameters: amplitude $\nu_k$, velocity $\mu_k$,
center of mass position $\xi_k$ and phase $\delta _k$.
The adiabatic approximation uses as a small parameter $\varepsilon_0
\ll 1$ the soliton overlap which falls off exponentially with the distance
between the solitons. Then the soliton parameters must satisfy
\cite{Karp}:
\begin{eqnarray}\label{eq:ad-ap}
&& |\nu _k-\nu _0| \ll \nu _0, \quad |\mu _k-\mu _0| \ll \mu _0,
\nonumber\\ && |\nu _k-\nu _0| |\xi_{k+1,0}-\xi_{k,0}| \gg 1,
\end{eqnarray}
where $\nu _0 = {1  \over N }\sum_{k=1}^{N}\nu _k$, and $ \mu _0 =
{1 \over N }\sum_{k=1}^{N}\mu _k$ are the average amplitude and
velocity respectively. In fact we have two different scales:
\[ |\nu _k-\nu _0| \simeq \varepsilon_0^{1/2}, \qquad
|\mu _k-\mu _0| \simeq \varepsilon_0^{1/2}, \]
\[ |\xi_{k+1,0}-\xi_{k,0}| \simeq \varepsilon_0^{-1}.
\]
One can expect that the approximation holds only for such times
$t$ for which the set of $4N$ parameters of the soliton train
satisfy (\ref{eq:ad-ap}).

Equation (\ref{eq:nls}) finds a number of applications to
nonlinear optics and for $R[u]\equiv 0 $ is integrable via the
inverse scattering transform method \cite{ZMNP,FaTa}. The $N
$-soliton train dynamics in the adiabatic approximation is
modelled by a complex generalization of the Toda chain
\cite{GKUE,GUED}:
\begin{equation}\label{eq:ctc}
\frac{{\rm d}^2Q_j}{{\rm  d}t^2}= 16\nu _0^2\left( e^{Q_{j+1}-Q_j}-
e^{Q_j-Q_{j-1}} \right) , \qquad j=1,\dots ,N.
\end{equation}
The complex-valued $Q_k $ are expressed through the soliton
parameters by:
\begin{equation}\label{hnls-Q}
Q_{k}(t) = 2i\lambda  _0 \xi_k(t) + 2k\ln (2\nu _0 ) +
i(k\pi -\delta_{k}(t)  -\delta_0 ),
\end{equation}
where $\delta _0=1/N \sum_{k=1}^{N}\delta _k $ and $\lambda _0=\mu _0+i\nu
_0 $. Besides we assume free-ends conditions, i.e., $e^{-Q_0}\equiv
e^{Q_{N+1}}\equiv 0$.

Note that the $N $-soliton train is {\em not} an $N $-soliton solution.
The spectral data of the corresponding Lax operator $L $ is nontrivial
also on the continuous spectrum of $L $. Therefore the analytical results
from the soliton theory can not be applied. Besides we want to treat
solitons moving with equal velocities and also to account for the effects
of possible nonintegrable perturbations $R[u] $.

The fact \cite{MFM,Moser} that the CTC, like the (real) Toda chain
(RTC) \cite{MFM}, is a completely integrable Hamiltonian system
allows one to analyze analytically the asymptotic behavior of the
$N$-soliton trains. However unlike the RTC, the CTC has richer
variety of dynamical regimes \cite{GKUE,GKDU,GEI} such as:
\begin{itemize}
\item asymptotically free motion if $v_j\neq v_k $ for $j\neq k $; this is
the only dynamical regime possible for RTC;

\item $N $-s bound state if $v_1 =\dots = v_N $ but
  $\zeta _k\neq \zeta _j $ for $k\neq j $;

\item  various intermediate (mixed) regimes; e.g., if $v_1= v_2 >
\dots > v_N $ but $\zeta _k\neq \zeta _j $ for $k\neq j $ then
we will have a bound state of the first two solitons while all the
others will be asymptotically free;

\item singular and degenerate regimes if two or more of the eigenvalues
of $L$ become equal, e.g., $\zeta _1=\zeta _2 \dots $ and $\zeta _j\neq
\zeta _k $ for $2<j\neq k $.
\end{itemize}

By $\zeta_k=v_k+iw_k$ above we have denoted the eigenvalues of the Lax
matrix $L$ in the Lax representation $L_\tau = [M,L]  $ of the CTC where:
\begin{eqnarray}\label{eq:L-M}
L&=& \sum_{k=1}^{N} b_k E_{kk} + \sum_{k=1}^{N-1} a_k ( E_{k,k+1} +
E_{k+1,k}),
\end{eqnarray}
where
\[ b_k \equiv  -{1 \over 2} {dQ_k  \over d\tau } = {\mu _k+i\nu_k\over 2 }
, \qquad  a_k = {1\over 2 } e^{(Q_{k+1}-Q_k)/2},
\]
and the matrices $E_{kp} $ are defined by
$(E_{kp})_{ij}=\delta_{ki}\delta_{pj}$.  The eigenvalues $\zeta _k $ of $L
$ are time independent and complex-valued along with the first components
$\eta _k=\vec{z}_1^{(k)}$ of the normalized eigenvectors of $L$:
\begin{equation}\label{eq:L-v}
L\vec{z}^{(k)} = \zeta _k \vec{z}^{(k)}, \qquad
(\vec{z}^{(k)},\vec{z}^{(m)})=\delta _{km}.
\end{equation}
The set of $\{\zeta_k = v_k +i w_k, \; \eta_k = \sigma_k +
i\theta_k\} $ may be viewed as the set of action-angle variables
of the CTC.

Using the CTC model  one can determine the asymptotic regime of
the $N $-soliton train.  Given the initial parameters  $\mu _k(0),
\nu _k(0), \xi_{k}(0), \delta _k(0) $ of the $N $-soliton train
one can calculate the matrix elements $b_k$ and $a_k$ of $L$ at
$t=0$. Then solving  the characteristic equation on $L|_{t=0}$ one
can calculate the eigenvalues $\zeta_k$ to  determine the
asymptotic regime of the $N $-soliton train \cite{GKUE,GKDU}.
Another option is to impose on $\zeta_k$ a specific constraint,
e.g. that all $\zeta _k $ be purely imaginary, i.e. all $v_k=0$.
This will provide a set of algebraic conditions  $L|_{t=0}$, and
on the initial soliton parameters $\mu _k(0), \nu _k(0),
\xi_{k}(0), \delta _k(0) $ which  characterize the region in the
soliton parameter space responsible for the $N$-soliton bound
states.

\section{The perturbed NLS and perturbed CTC equations}
\label{sec:Pert}

We will consider several specific choices $R^{(p)}[u] $ of
perturbations, $p=1,2,\dots $ in (\ref{eq:nls}). In the adiabatic
approximation the dynamics of the soliton parameters can be
determined by the system (see \cite{Karp} for $N=2 $ and
\cite{GKUE,GKDU} for $N>2 $):
\begin{eqnarray}\label{eq:nuk0}
{d\lambda _k \over dt} &=& -4\nu _0 \left(e^{Q_{k+1}-Q_k} -
e^{Q_k -Q_{k-1}} \right) \nonumber\\
&+&  M^{(p)}_k +i N^{(p)}_k , \\
\label{eq:xik0}
{d \xi_k \over dt} &=& 2 \mu_k +   \Xi^{(p)}_k  ,\qquad
{d \delta_k \over dt} = 2 (\mu_k^2 + \nu_k^2)+   X_k^{(p)} ,
\end{eqnarray}
where $\lambda _k=\mu _k+i\nu _k $ and $X_k^{(p)}  = 2 \mu_k \Xi_k^{(p)}
+ D_k^{(p)} $.  The right hand sides of Eqs.
(\ref{eq:nuk0})--(\ref{eq:xik0}) are determined by $R_k^{(p)}[u]$ through:
\begin{eqnarray}\label{eq:Nk0}
N_k^{(p)}  &=& {1 \over 2} \int_{-\infty}^{\infty} {dz_k \over \cosh z_k
}\, \mbox{Re}\,  \left( R_k^{(p)} [u] e^{-i\phi_k} \right) ,\\
\label{eq:Mk0}
M_k^{(p)}  &=& {1 \over 2} \int_{-\infty}^{\infty} {dz_k  \, \sinh z_k
\over \cosh^2 z_k }\,\mbox{Im}\,\left( R_k^{(p)} [u] e^{-i\phi_k} \right),
\\
\label{eq:Xik0}
\Xi_k^{(p)}  &=& {1 \over 4 \nu_k^2} \int_{-\infty}^{\infty} { dz_k \,
z_k\over \cosh z_k }\, \mbox{Re}\, \left( R_k^{(p)} [u] e^{-i\phi_k}
\right),
\end{eqnarray}
\begin{equation}\label{eq:Dk0}
D_k^{(p)}  = {1 \over 2 \nu_k} \int_{-\infty}^{\infty} {dz_k \,
( 1 - z_k \tanh z_k)  \over \cosh z_k }
\mbox{Im}\, \left( R_k^{(p)} [u] e^{-i\phi_k} \right) .
\end{equation}

Inserting (\ref{eq:nuk0}), (\ref{eq:xik0}) into (\ref{hnls-Q}) we derive:
\begin{eqnarray}
{dQ_k  \over dt } &=& -4\nu _0\lambda _k + {2k \over\nu _0 }
\mathcal{N}_{0}^{(p)} + 2i\xi_k   \left( \mathcal{M}_{0}^{(p)} +
i \mathcal{N}_{0}^{(p)}\right) \nonumber\\
\label{eq:pctc}
&+& i   \left( 2\lambda _0 \Xi_{k}^{(p)} - X_{k}^{(p)} -
\mathcal{X}_{0}^{(p)}\right),\\
\mathcal{N}_{0}^{(p)} &=& {1\over N} \sum_{j=1}^{N} N_{j}^{(p)}, \qquad
\mathcal{M}_{0}^{(p)} = {1\over N} \sum_{j=1}^{N} M_{j}^{(p)}, \nonumber\\
\mathcal{X}_{0}^{(p)} &=& {1\over N} \sum_{j=1}^{N} X_{j}^{(p)}.
\nonumber
\end{eqnarray}

In deriving eq. (\ref{eq:pctc}) we have kept terms of the order
$\Delta \nu _k \simeq \mathcal{O}(\sqrt{\epsilon_0 }) $ and
neglected terms of the order $\mathcal{O}(\epsilon_0 ) $. The
perturbations result in that $\nu _0 $ and $\mu _0 $ may become
time-dependent. Indeed, from (\ref{eq:nuk0}) we get:
\begin{equation}\label{eq:lam0}
{d\mu _0 \over dt } = \mathcal{M}_{0}^{(p)} , \qquad
{d\nu _0 \over dt } = \mathcal{N}_{0}^{(p)} ,
\end{equation}

The small parameter $\epsilon _0 $ can be related to the initial
distance $r_0=|\xi_2 -\xi_1|_{t=0} $ between the two solitons. Assuming
$\nu _{1,2}\simeq \nu _0 $ we find:
\begin{equation}\label{eq:eps0}
\epsilon _0 = \int_{-\infty }^{\infty } dx \left| u_{1}^{\rm 1s}(x,0)
u_{2}^{\rm 1s}(x,0)\right| \simeq 8\nu _0 r_0e^{-2\nu _0r_0}.
\end{equation}
In particular, (\ref{eq:eps0}) means that $\epsilon _0\simeq 0.01 $ for
$r_0\simeq 8 $ and $\nu _0=1/2 $.

We assume that initially the solitons are ordered in such a way that
$\xi_{k+1}-\xi_k \simeq r_0$.  One can check \cite{GUED,GU} that
$N_k^{(p)}\simeq M_k^{(p)} \simeq \exp (- 2\nu _0|k-p| r_0) $. Therefore
the interaction terms between the $k $-th and $k\pm 1 $-st solitons will
be of the order of $e^{-2\nu _0r_0} $; the interactions between $k $-th
and $k\pm 2 $-nd soliton will of the order of $e^{-4\nu_0r_0} \ll
e^{-2\nu_0r_0}$.

The terms $\Xi_{k}^{(0)} $, $X_{k}^{(0)} $ are of the order of
$r_0^a\exp(-2\nu_0r_0) $, where $a=0 $ or $1 $. However they can be
neglected as compared to $\widetilde{\mu }_k $ and $\widetilde{\nu }_k $,
where
\begin{equation}\label{eq:mu_k}
\widetilde{\mu }_k = \mu _k-\mu _0 \simeq \sqrt{\epsilon_0 }, \qquad
\widetilde{\nu }_k = \nu _k-\nu _0\simeq \sqrt{\epsilon_0 },
\end{equation}

The corrections to $N_{k}^{(p)} $, \dots, coming from the terms
linear in $u $ depend only on the parameters of the $k $-th
soliton; i.e., they are `local' in $k $. The nonlinear terms in $u
$ present in $iR^{(p)}[u] $ produce also `non-local' in $k $ terms
in $N_{k}^{(p)} $, \dots.

\subsection{Quadratic and tilted potentials}\label{Quadratic}

Let $iR[u]=V(x)u(x,t) $. Our first choice for $V(x) $ is a quadratic one:
\begin{equation}\label{eq:P3.5}
V^{(1)}(x) = V_2x^2 + V_1x + V_0.
\end{equation}
Skipping the details we get the results:
\begin{subequations}\label{eq:P3.4}
\begin{eqnarray}\label{eq:P3.4a}
N_{k}^{\rm (1)}  &=& 0, \qquad \Xi_{k}^{\rm (1)}  = 0 , \\
M_{k}^{\rm (1)}  &=& -V_2 \xi_k - {V_1  \over 2 } , \\
\label{eq:P3.4c}
D_{k}^{\rm (1)}  &=& V_2\left( {\pi^2  \over 48\nu _k^2 } -
\xi_k^2\right) -V_1 \xi_k -V_0,
\end{eqnarray}
\end{subequations}
and $X_{k}^{\rm  (1)}  = D_{k}^{\rm (1)}  $. As a result the corresponding
PCTC takes the form \cite{Liss}:
\begin{eqnarray}\label{eq:P4.3a}
{d\lambda _k \over dt } &=& -4\nu _0 \left( e^{Q_{k+1}-Q_{k}}
-  e^{Q_{k}-Q_{k-1}} \right) \nonumber\\
&-& V_2\xi_k - {V_1  \over 2 }, \\
\label{eq:P4.3b}
{dQ_k  \over dt } &=& - 4\nu _0(\mu _k+i\nu _k) - iD_{k}^{\rm (1)}
- {i \over N } \sum_{j=1}^{N} D_{j}^{\rm (1)} ,
\end{eqnarray}
where $\lambda _k=\mu _k+i\nu _k $.

If we now differentiate (\ref{eq:P4.3b}) and make use of (\ref{eq:P4.3a})
we get \cite{Liss}:
\begin{eqnarray}\label{eq:P4.4}
{d^2 Q_k  \over dt^2 } &=& 16\nu _0^2 \left( e^{Q_{k+1}-Q_{k}}
-  e^{Q_{k}-Q_{k-1}} \right) \\
&+& 4\nu _0 \left(V_2\xi _k + {V_1  \over
2}\right) - i{d D_{k}^{\rm (1)}  \over dt }- { i \over N }
\sum_{j=1}^{N} {d D_{j}^{\rm (1)}  \over dt}. \nonumber
\end{eqnarray}

It is reasonable to assume that $V_2\simeq {\cal  O}(\epsilon_0/N
)$; this ensures the possibility to have the $N $-soliton train
`inside' the potential. It also means that both the exponential
terms and the correction terms $M_{k}^{\rm (1)}  $ are of the same
order of magnitude. From eqs. (\ref{eq:P4.3a}) and
(\ref{eq:P4.3b}) there follows that $d\nu _0/dt =0 $ and:
\begin{equation}\label{eq:mu0-xi0}
{d\mu _0 \over dt } = -V_2 \xi_0 -{V_1 \over 2 }, \qquad
{d\xi_0  \over dt } = 2\mu _0,
\end{equation}
where $\mu _0 $ is the average velocity and $ \xi _{0} = {1 \over N }
\sum_{j=1}^{N} \xi _j$, is the center of mass of the $N $-soliton train.
The system of equations (\ref{eq:mu0-xi0}) for $V_2>0 $ has a simple
solution
\begin{eqnarray}\label{eq:mu0-xi0s}
\mu _0(t) &=& \mu _{00} \cos (\Phi (t)), \nonumber\\
\xi _0(t) &=& \sqrt{ 2 \over
V_2 } \mu _{00} \sin (\Phi (t)) - {V_1  \over 2V_2 },
\end{eqnarray}
where $\Phi (t)=\sqrt{2V_2}t +\Phi _0 $, and $\mu _{00} $ and $\Phi _0 $
are constants of integration. Therefore the overall effect of such
quadratic potential will be to induce a slow periodic motion of the train
as a whole.

By tilted potential below we mean a particular case of the quadratic
potential with $V_2=0 $. Then the equations
(\ref{eq:mu0-xi0})-(\ref{eq:mu0-xi0s}) take the form:
\begin{equation}\label{eq:mu0-xi0t}
{d\widetilde{\mu} _0 \over dt } = -{V_1 \over 2 }, \qquad
{d\widetilde{\xi}_0  \over dt } = 2\widetilde{\mu} _0,
\end{equation}
and have the  simple solution
\begin{eqnarray}\label{eq:mu0-xi0st}
\widetilde{\mu} _0(t) &=&  -{V_1 \over 2 } t +\widetilde{\mu }_{00},
\nonumber\\
\widetilde{\xi} _0(t) &=&  -{V_1 \over 4 }t^2 +\widetilde{\mu }_{00}t
+\widetilde{\xi }_{00},
\end{eqnarray}
Therefore the tilted potential accelerates the soliton train
as a whole in a prescribed direction. It can be used also to `pick up' and
accelerate one or more of the solitons from the train confined to a periodic
potential.

\subsection{Periodic potentials}\label{Periodic}

One may consider several physically important choices of periodic
potentials. The simplest one is
\begin{eqnarray}\label{eq:V-per}
V^{(2)}(x) &=& A_1 \cos(\Omega_1 x+\Omega _0) \nonumber\\
&=& A_1 - 2A_1 \sin^2 (\Omega_1 x/2 +\Omega _0/2),
\end{eqnarray}
where $A $, $\Omega  $ and $\Omega _0 $ are appropriately chosen
constants. NLS equation with similar potentials appear in a natural way in
the study of Bose-Einstein condensates, see \cite{BCCD}.

The relevant integrals for $N_k $, $M_k $, $\Xi_k $ and $D_k $ are equal
to \cite{Liss}:
\begin{eqnarray}\label{eq:P16.1}
N_k^{(2)}&=&0, \qquad \Xi_k^{(2)}=0, \\
M_k^{(2)}&=&{\pi A_1\Omega_1^2 \over 8\nu _k }
{1 \over \sinh Z_k }\sin (\Omega_1 \xi_k +\Omega _0), \\
\label{eq:P16.2}
D_k^{(2)}&=& -{\pi^2 A_1\Omega_1 ^2 \over 16\nu_k^2 }
{\cosh Z_k  \over \sinh ^2 Z_k } \cos (\Omega_1 \xi_k +\Omega _0) ,
\end{eqnarray}
where $Z_{1,k} =\pi\Omega_1 /(4\nu _k) $. These results allow one to
derive the corresponding perturbed CTC models. Again we find that $d\nu
_0/dt =0 $.

The second case we will consider is a linear combination of two potentials
of the form (\ref{eq:V-per}) with correlated frequences:
\begin{equation}\label{eq:V-per3}
V^{(3)}(x) = A_1 \cos(\Omega_1 x+\Omega _0)
+ A_2 \cos(\Omega_2 x+\Omega _0)
\end{equation}
where $A_j $, $\Omega_j  $ and $\Omega _0 $ are appropriately chosen
constants. Such potentials with correlated frequences, e.g. $\Omega
_2=2\Omega _1 $ also appear in the study of Bose-Einstein
condensates, see \cite{BCCD}.

The relevant integrals for $N_k $, $M_k $, $\Xi_k $ and $D_k $ are equal
to \cite{Liss}:
\begin{eqnarray}\label{eq:P16.1-3}
N_k^{(3)}&=&0, \qquad \Xi_k^{(3)}=0, \\
M_k^{(3)} &=&
A_1M_{k}(\Omega _1,Z_{1,k})\sin (\Omega_1 \xi_k +\Omega _0)
\nonumber\\ &+& A_2M_{k}(\Omega _2,Z_{2,k})\sin (\Omega_2 \xi_k
+\Omega _0), \\
\label{eq:P16.2-3}
D_k^{(3)} &=& A_1D_{k}(\Omega _1,Z_{1,k})\cos (\Omega_1 \xi_k
+\Omega _0) \nonumber\\ &+& A_2D_{k}(\Omega _2,Z_{2,k})\cos
(\Omega_2 \xi_k +\Omega _0),
\end{eqnarray}
where
\begin{eqnarray}\label{eq:M-Djk}
M_{k}(\Omega _j,Z_{j,k}) &=& {\pi \Omega_j^2 \over 8\nu _k }
{1 \over \sinh Z_{j,k} }, \nonumber\\
D_{k}(\Omega _j,Z_{j,k})&=&- {\pi^2 \Omega_j ^2 \over
16\nu_k^2 } {\cosh Z_{j,k}  \over \sinh ^2 Z_{j,k} } ,
\end{eqnarray}
$Z_{j,k} =\pi\Omega_j /(4\nu _k) $. These results
allow one to derive the corresponding perturbed CTC models. Again we find
that $d\nu _0/dt =0 $.

The last case we consider is an elliptic potential of the form:
\begin{eqnarray}\label{eq:V-per4}
&&V^{(4)}(x) = -B \sn^2(\Omega x;k)\\
&&{}=-B\sum_{s=0}^{\infty } G_s(k) \sin (\Omega _s x), \nonumber\\
\label{eq:Gs} &&G_s(k) = \left( {1+k^2 \over 2k^3} - {(2s+1)^3
\pi^2\over 8k^3K^2}
\right) {\pi\over K\sinh(\omega_s ) },\\
&&\Omega _s= {(2s+1)\pi \over 2K }\Omega , \qquad \omega_s = {(2s+1)
\pi K' \over 2K},
\end{eqnarray}
where $k $ is the module of the elliptic function, $K=K(k)$ is the
complete elliptic integral of the first kind, and $B $ is a
constant.

The second line of formula (\ref{eq:V-per4}) provides the expansion of
$\sn^2(\Omega x;k) $ as infinite series of trigonometric functions. Each
of the terms in this series can be treated as perturbation just like
above assuming $\Omega _0=-\pi/2 $. As a result we get:
\begin{equation}\label{eq:P16.1-4}
N_k^{(4)}=0, \qquad \Xi_k^{(4)}=0,
\end{equation}
\begin{equation}\label{eq:Mk-4}
M_k^{(4)} = B\sum_{s=0}^{\infty }
M_{k}(\Omega _s,Z_{s,k})G_s(k)\cos (\Omega_s \xi_k ),
\end{equation}
\begin{equation}\label{eq:P16.2-4}
D_k^{(4)} = -B\sum_{s=0}^{\infty }D_{k}(\Omega _s,Z_{s,k})G_s(k)
\sin(\Omega_s \xi_k ) ,
\end{equation}
where $M_{k} $, $D_{k}$ and  $\Omega _s=(2s+1)\Omega  $ are defined as
in (\ref{eq:M-Djk}).

\section{CTC analysis and comparison with
         numerical simulations} \label{sec:Num}

The dynamics of an individual soliton in a train is determined by
the combined action of external potential and the influence of
neighboring solitons. The interaction with neighboring solitons
can be either repulsive, or attractive depending on the phase
relations between them. Particularly, if their amplitudes are
equal and the initial phase difference between neighboring
solitons is $\pi$ (as considered below) they repel each other
giving rise to expanding motion in the absence of an external
field \cite{GKUE,GUED}.

The external potential counterbalances the expansion, trying to
confine solitons in the minima of the potential. It is the
interplay of these two factors - the interaction of solitons and
the action of the external potential, which gives rise to a rich
dynamics of the $N$-soliton train.

To verify the adequacy of the perturbed CTC model for the description
of the $N$-soliton train dynamics in external potentials we
performed comparison of predictions of corresponding perturbed CTC
(PCTC) system and direct simulations of the underlying NLS
equation (\ref{eq:nls}). Below we present results pertaining to a
matter-wave soliton train in a confining (i) parabolic trap and in
(ii) a periodic potential modelling an optical lattice.

Here we present the numerical verification of the PCTC model.
The perturbed NLS eq. (\ref{eq:nls}) is solved by the operator splitting
procedure using the fast Fourier transform \cite{taha}. In the course of
time evolution we monitor the conservation of the norm and energy of
the N-soliton train. The corresponding PCTC equations are solved by the
Runge-Kutta scheme with the adaptive stepsize control \cite{numrecipes}.

The evolution of a $N$-soliton train in the absence of potential
($V(x)=0$) is well known, see e.g. \cite{GUED,GKDU}. These papers propose
a method to determine the asymptotic dynamical regime of the CTC for a
given set of initial parameters $\nu _k(0) $, $\mu _k(0) $, $\xi _k(0) $
and $\delta _k(0) $. Below we will use mainly the following set of
parameters $\nu _k(0) =1/2$, $\mu _k(0) =0$, $\xi_{k+1}(0) -\xi_k(0)= r_0$
with two different choices for the phases:
\begin{eqnarray}\label{eq:in-par}
\delta _k(0) =k\pi,\\
\label{eq:in-para}
\delta _k(0) =0.
\end{eqnarray}
These two types of initial conditions (IC) are most widely used in numeric
simulations.

In the absence of potential the IC (\ref{eq:in-par}) ensure the so called
free  asymptotic regime, i.e.  each soliton develops its own velocity and
the distance between the neighboring solitons increases linearly in time.
At the same time the center of mass of the soliton train stays at rest
(see the left panel of Fig.~\ref{fig1}).  Under the IC (\ref{eq:in-para})
the solitons attract each other going into collisions whenever the
distance between them is not large enough.

From mathematical point of view the IC (\ref{eq:in-par}) reduce
the CTC into a standard (real) Toda chain for which the free
asymptotic regime is the only possible asymptotical regime. On the
contrary the IC (\ref{eq:in-para}) lead to singular solutions for
the CTC (see \cite{GKUE,GKDU,GEI}). The singularities of the exact
solutions for the CTC coincide with the positions of the
collisions.

Below we will study the effects of the potentials for both types of IC.
One may expect that the quadratic potential will prevent free asymptotic
regime of IC (\ref{eq:in-par}) no matter how small $V_2>0 $ is and would
not be able to prevent collisions in the case of IC (\ref{eq:in-para}).
The periodic potential, if strong enough, should be able to stabilize and
bring to bound states both types of IC.

\subsection{Quadratic potential}

For the quadratic external potential $V(x) = V_2 x^2 + V_1 x + V_0$ the
perturbed CTC equations in terms of soliton parameters have the form:
\begin{eqnarray}\label{eq:mu-k}
{d\mu _k \over dt } &=& 16\nu _0^3 \left( e^{-2\nu _0 (\xi_{k+1}
-\xi_{k})} \cos \Phi _k \right.  \nonumber\\
&-& \left. e^{-2\nu _0 (\xi_{k}
-\xi_{k-1})} \cos \Phi _{k-1}\right) -V_2\xi_k - {V_1 \over 2 },\\
\label{eq:nu-k} {d\nu _k \over dt } &=& 16\nu _0^3 \left(
e^{-2\nu _0 (\xi_{k+1} -\xi_{k})} \sin \Phi _k \right.  \nonumber\\
&&\qquad - \left. e^{-2\nu _0 (\xi_{k} -\xi_{k-1})} \sin \Phi
_{k-1}\right),\\
\label{eq:xi-k} {d\xi _k \over dt } &=& 2\mu _k,
\end{eqnarray}
\begin{equation}\label{eq:del-k}
{d\delta _k \over dt } = 2(\mu _k^2 +\nu _k^2) + V_2 \left( {\pi^2
\over 48\nu _k^2} -\xi_k^2\right) -V_1\xi_k -V_0 ,
\end{equation}
\begin{equation}
\label{eq:Phi-k}
\Phi _k= 2\mu _0(\xi_{k+1} -\xi_{k}) +\delta _{k} -\delta _{k+1} ,
\end{equation}
where $\mu _k $, $\nu _k $, $\xi_k $ and $\delta _k $ for $k=1,\dots,N $
are the $4N $ soliton parameters, see eqs.
(\ref{eq:IC})-(\ref{eq:nls-1s}).

\begin{figure}

\includegraphics[width=3.7cm,height=5cm,clip]{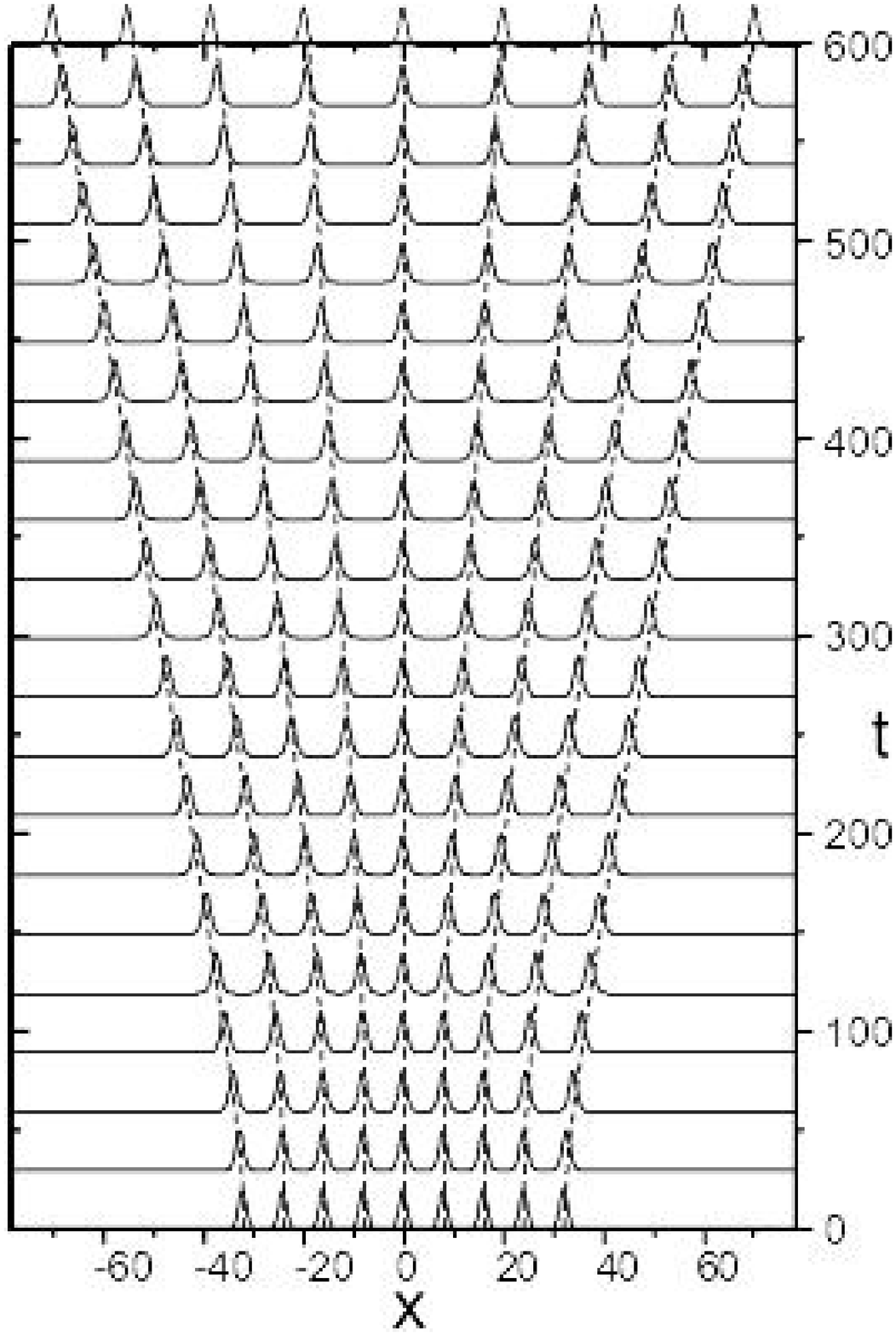}\qquad
\includegraphics[width=3.7cm,height=5cm,clip]{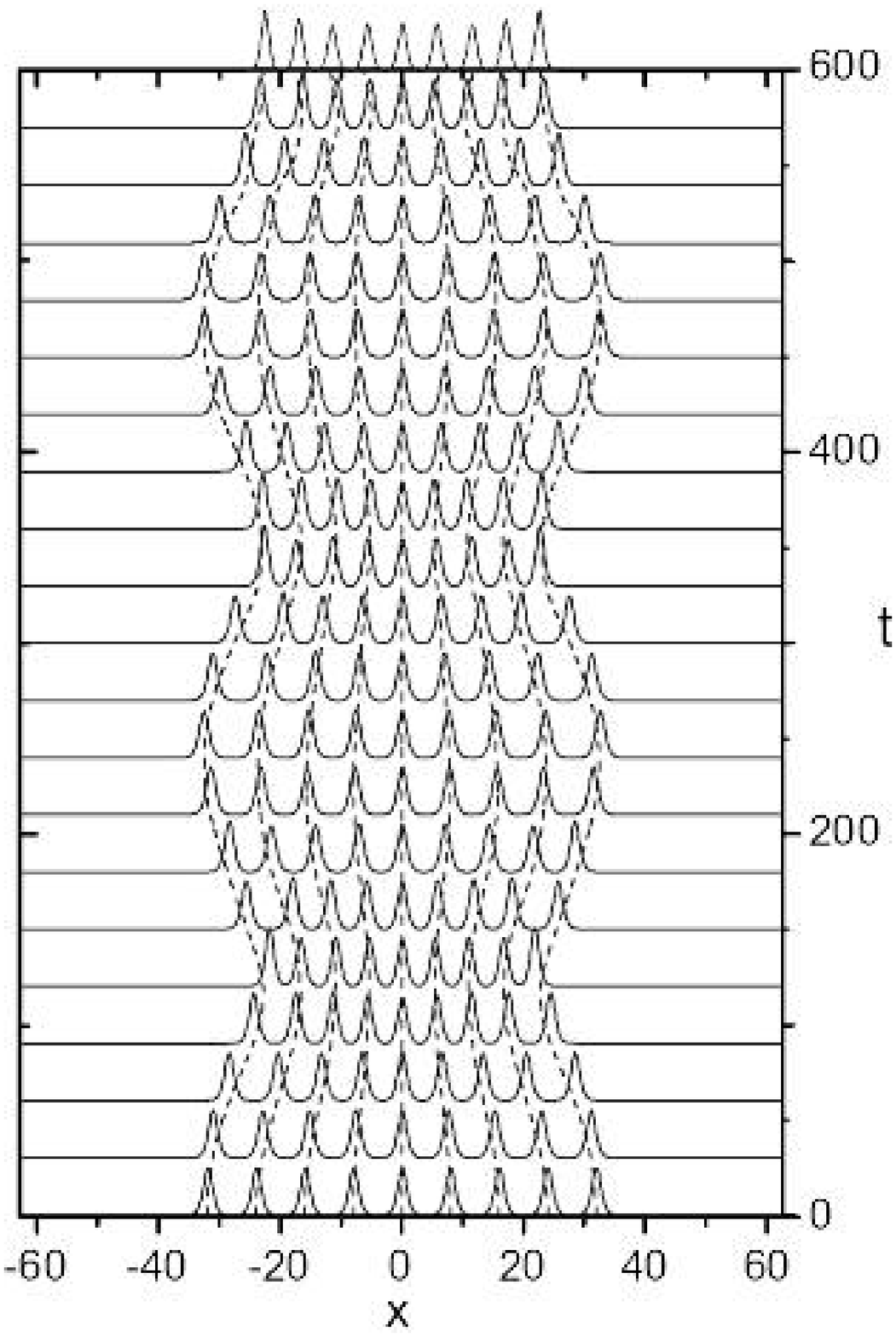}
\caption{Left panel: $9$-soliton train with initial parameters as
in eq. (\ref{eq:in-par}) with $r_0=8 $ in the absence of a
potential  goes into free asymptotic regime.  Solid lines:  direct
numerical simulation of the NLS equation (\ref{eq:nls});  dashed
lines:  $\xi_k(t) $ as predicted by the CTC equations
(\ref{eq:mu-k})-(\ref{eq:del-k}) with $V_0 = V_1 = V_2 = 0$ and
$r_0=8 $.  Right panel:  Evolution of a $9$-soliton train with the
same initial parameters in the quadratic potential $V(x) = V_2x^2$
with $V_2=0.00005 $. Solid lines: direct numerical simulation  of
the NLS equation (\ref{eq:nls}); dashed lines: solution of the
PCTC equations (\ref{eq:mu-k})-(\ref{eq:del-k}). Initially the
train is placed symmetrically relative to the minimum of the
potential at $x=0$. } \label{fig1}
\end{figure}

\begin{figure}
\includegraphics[width=3.7cm,height=5cm,clip]{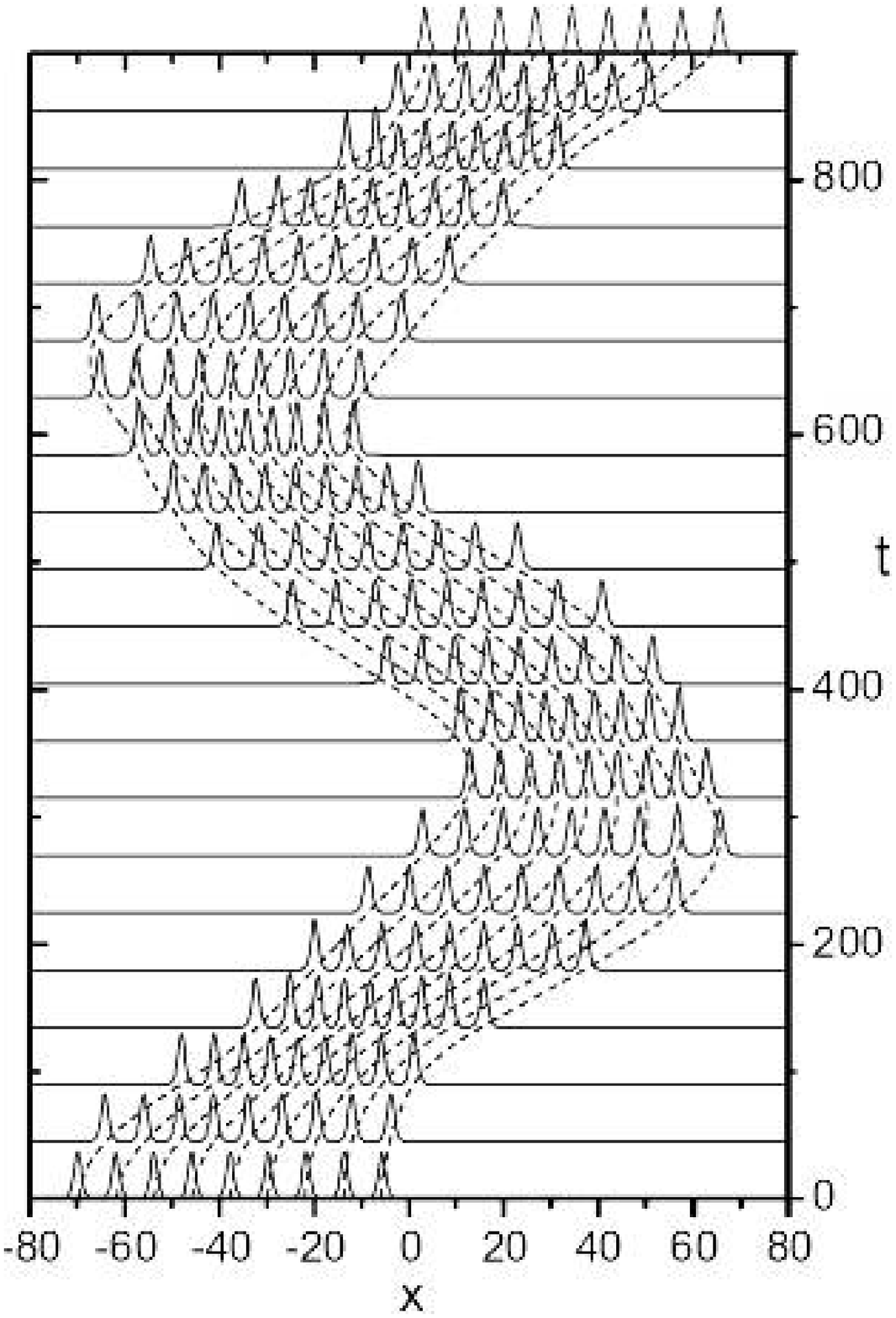} \qquad
\includegraphics[width=3.7cm,height=5cm,clip]{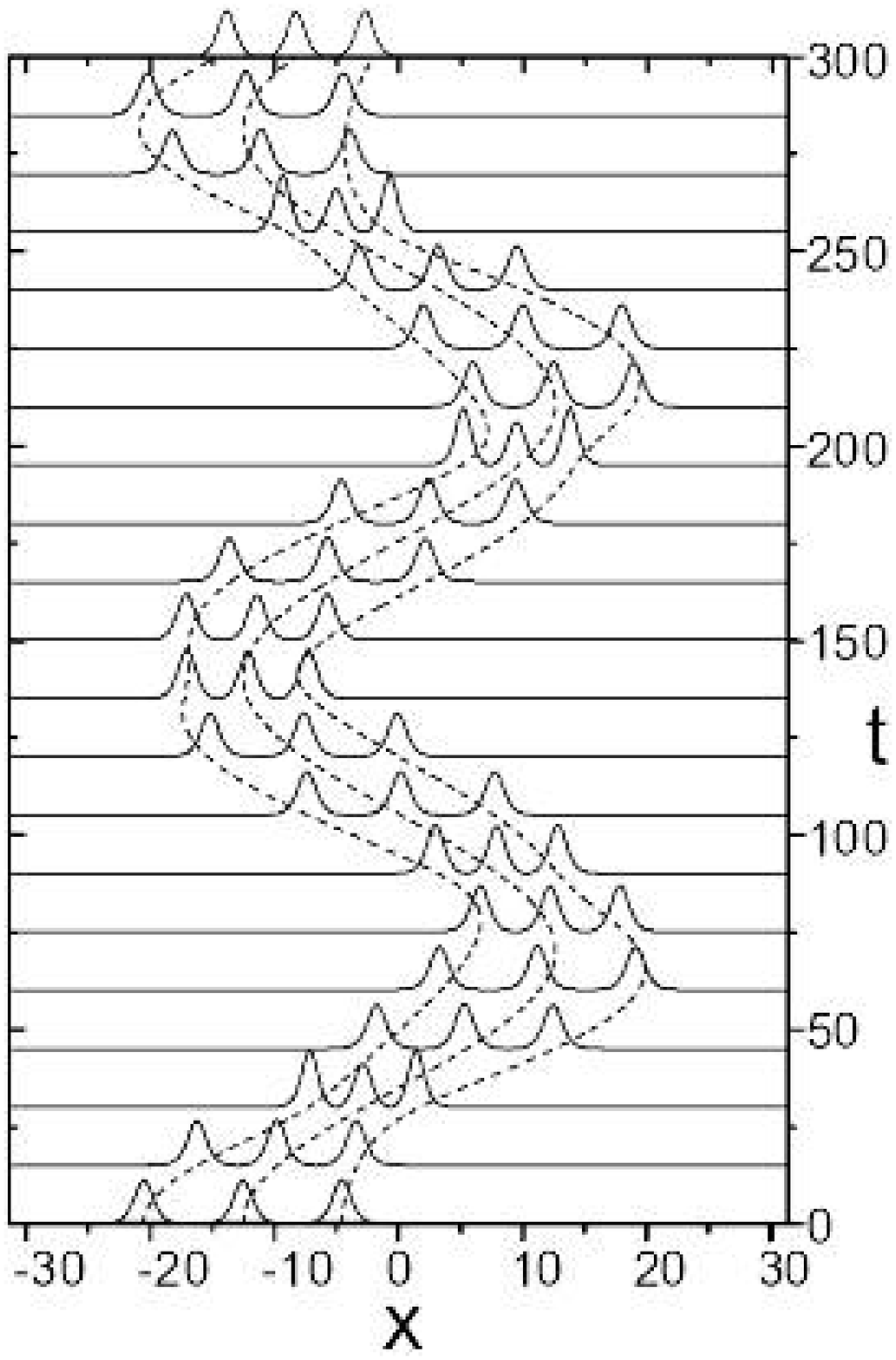}
\caption{Harmonic oscillations of a N-soliton train initially
shifted relative to the minimum of the quadratic potential $V(x) =
V_2x^2$. Left panel: 9-soliton train, $V_2 = 0.00005$. Right
panel: 3-soliton train, $V_2 = 0.001$. The IC of the both trains
is given by (\ref{eq:in-par}) with $r_0=8 $. In both panels solid
lines correspond to direct simulations of the NLS equation
(\ref{eq:nls}), and dashed lines to numerical solution of the PCTC
equations (\ref{eq:mu-k}) - (\ref{eq:del-k}).}
\label{fig2}
\end{figure}

The effect of the quadratic potential on the $N $-soliton train
with parameters (\ref{eq:in-par}) is to balance the repulsive
interaction between the solitons, so that they remain bounded by
the potential, as illustrated in  the right panels of the figures
\ref{fig1} and \ref{fig2}. The quadratic potentials are supposed
to be weak, i.e. we choose $V_2 $ so that
\begin{equation}\label{eq:v2}
V_2 \xi_N^2(0) \leq \nu _0, \qquad V_2 \xi_1^2(0) \leq \nu _0.
\end{equation}

\begin{figure}
\includegraphics[width=7cm,height=5cm,clip]{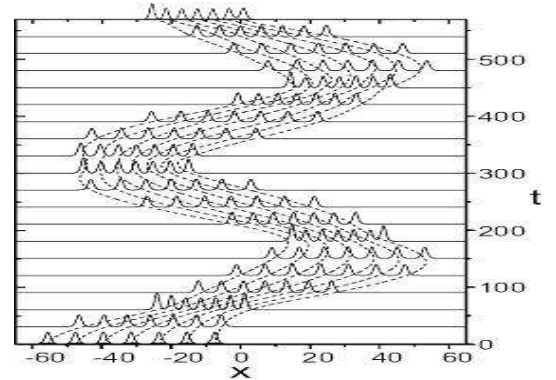}
\caption{Dynamics of a 7-soliton train placed asymmetrically
relative to the minimum of the trap $V(x) = 0.0002 x^2$. Solid
lines: results of direct numerical simulations of the NLS equation
(\ref{eq:nls}). Dashed lines: result of solution of the PCTC
system (\ref{eq:mu-k}) - (\ref{eq:del-k}) for the center of mass
$\xi_i$. The parameters of solitons are the same as in
(\ref{eq:in-par}) with $r_0=8$. The initial shift of the soliton
train relative to the minimum of the parabolic trap is $10 \pi$. }
\label{fig3}
\end{figure}

Figures \ref{fig1} and \ref{fig2} show good agreement between the PCTC
model and the numerical solution of the perturbed NLS equation
(\ref{eq:nls}).  They also show two types of effects of the quadratic
potential on the motion of the $N $-soliton train: (i) the train performs
contracting and expanding oscillations if its center of mass coincides
with the minimum of the potential, (ii) the train oscillates around the
minimum of the potential as a whole if its center of mass is shifted. In
the last case contracting and expanding motions of the soliton train is
superimposed to the center of mass dynamics.  As one can see from the
figures the period of this motion matches very well the one predicted by
formula (\ref{eq:mu0-xi0s}). Indeed, from eq. (\ref{eq:mu0-xi0s}) it
follows that the period of the center of mass motion is
$T=2\pi/\sqrt{2 V_2} $.  For the parameters in fig.~\ref{fig2} we have $T
\simeq 628$ (for 9-soliton train), $T \simeq 140$ (for $3$-soliton train).
Similar is the dynamics also for the $7 $-soliton train on
Fig.~\ref{fig3};  for the parameters choosen there we have $T \simeq 314$,
in good agreement with the numerical simulations.
The direct simulations of the NLS equation (\ref{eq:nls}) shows
that stronger parabolic trap may cause merging of individual
solitons at times of contraction, and restoring of the original
configuration when the train is expanded. This behavior reminds
the phenomenon of "missing solitons" observed in the experiment
\cite{strecker}. However, this situation is beyond the validity of
the PCTC approach.

\subsection{Tilted periodic potential}

Now we consider the dynamics of a N - soliton train in a tilted periodic
potential, which is the combination of periodic and linear potentials
\begin{equation} \label{tilted}
V(x) = A \cos(\Omega x + \Omega_0) + B x.
\end{equation}
This potential is of particular interest in studies of
Bose-Einstein condensates. A train of repulsive BEC loaded in such
a potential (where the periodic potential was a 1D optical lattice
and the linear one was due to the gravitation) exhibited Bloch
oscillations \cite{anderson}. At each period of these oscillations
condensate atoms residing in individual optical lattice cells
coherently tunneled through the potential barriers. This was the
first experimental demonstration of a pulsed atomic laser
\cite{anderson}. Recently a new model of a pulsed atomic laser was
theoretically developed in \cite{carr}, where the solitons of
attractive BEC were considered as carriers of coherent atomic
pulses.

\begin{figure}
{\includegraphics[width=3.7cm,height=5cm,clip]{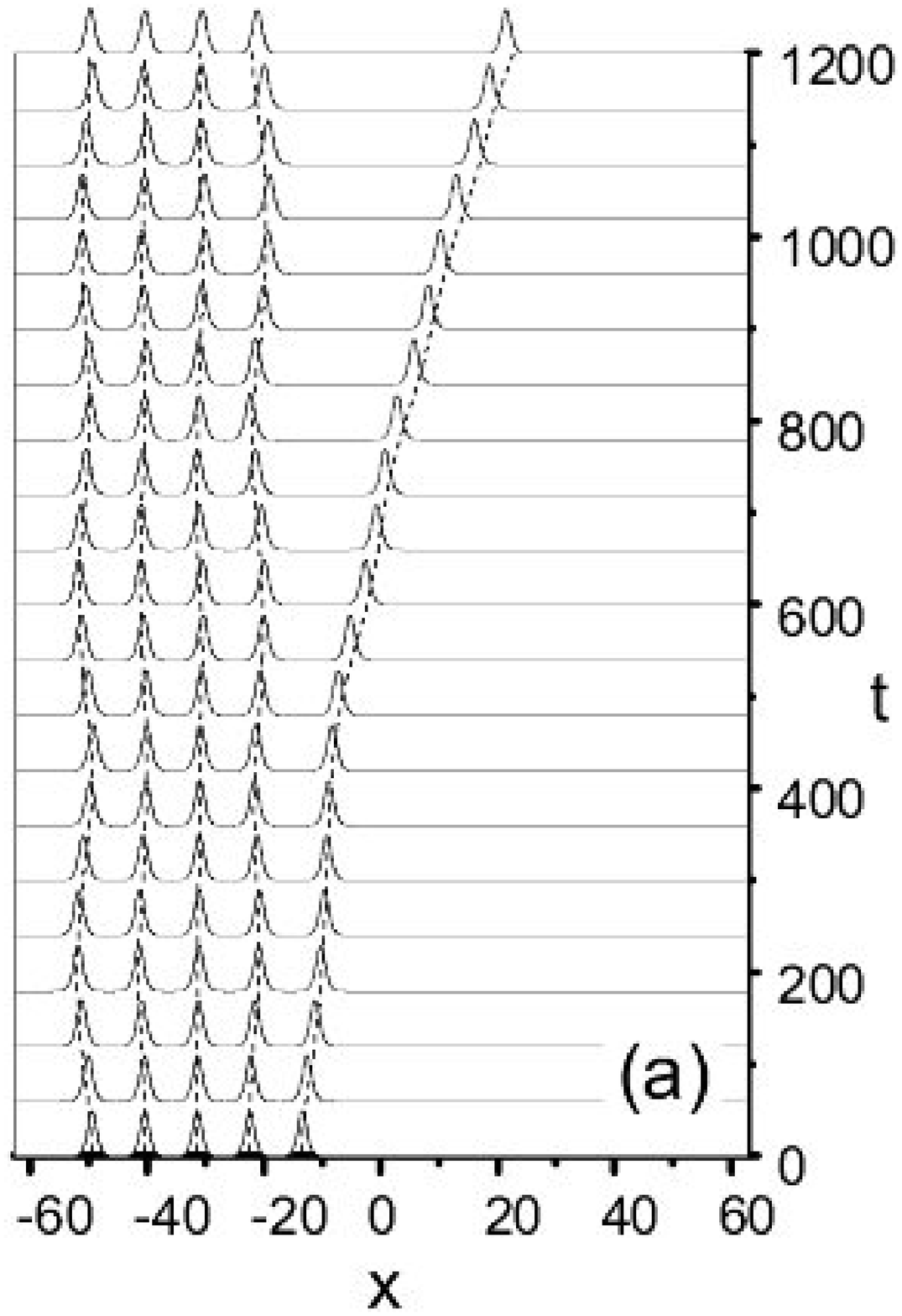} \qquad
\includegraphics[width=3.7cm,height=5cm,clip]{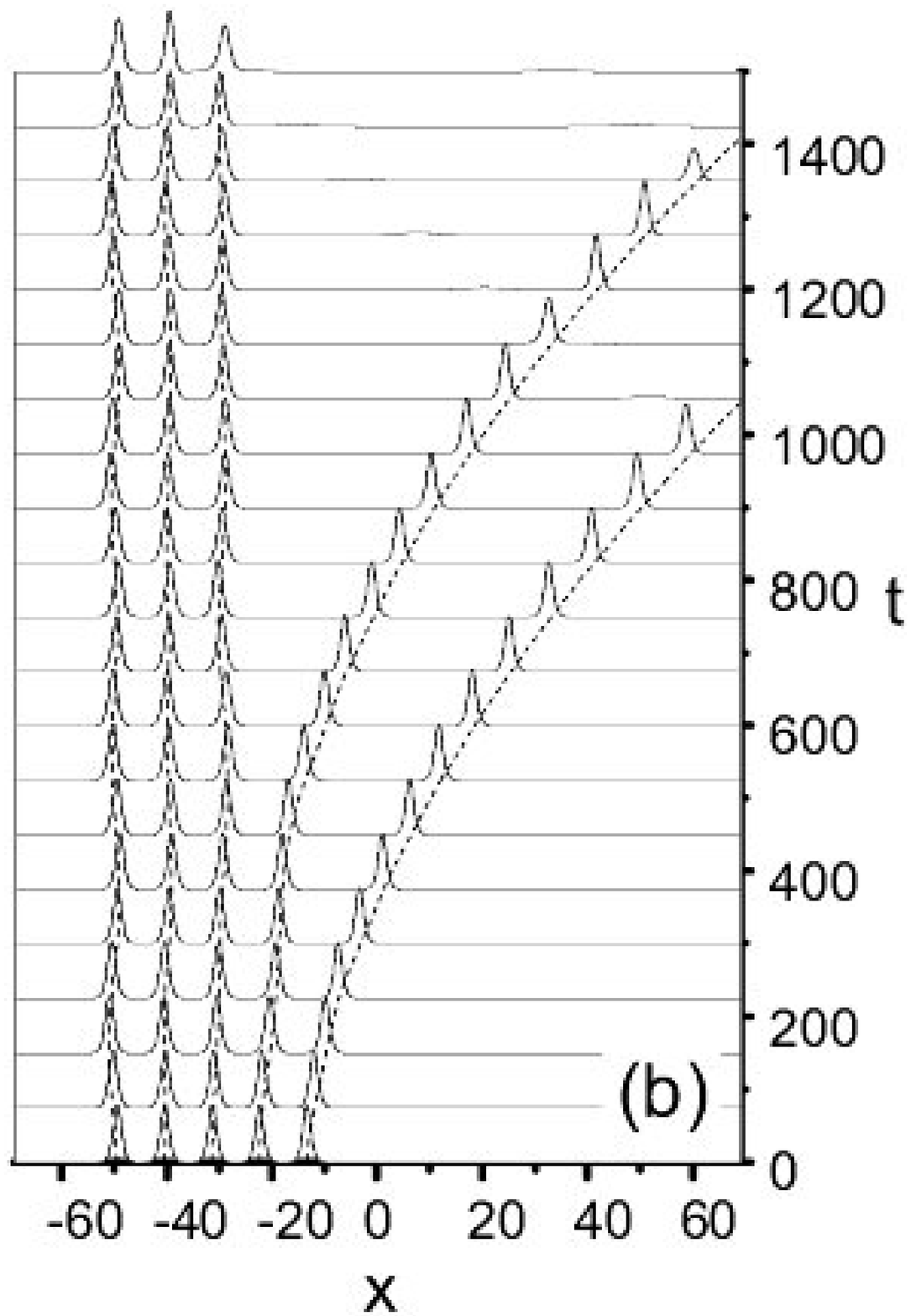}}\\
{\includegraphics[width=3.7cm,height=5cm,clip]{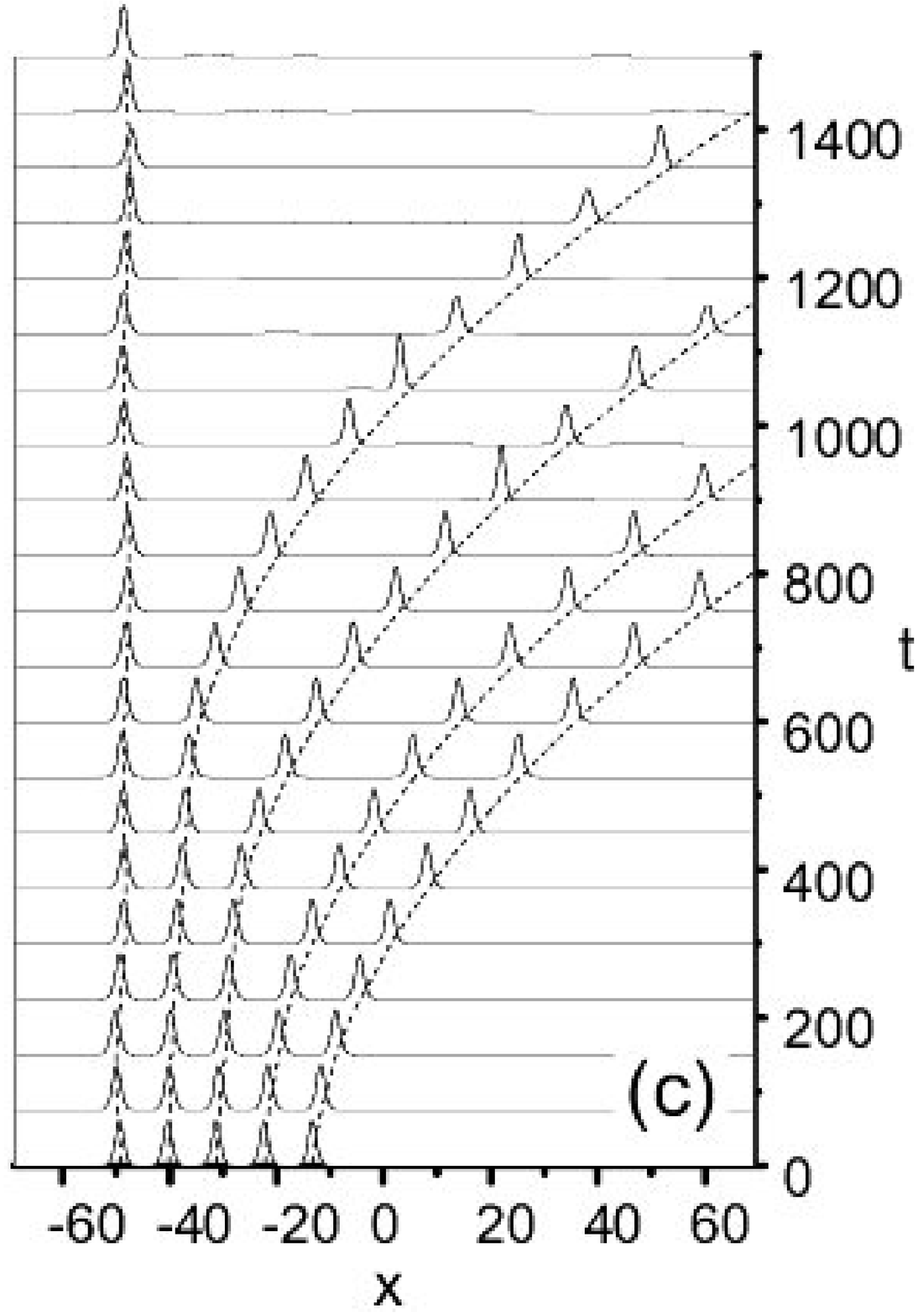} \qquad
\includegraphics[width=3.7cm,height=5cm,clip]{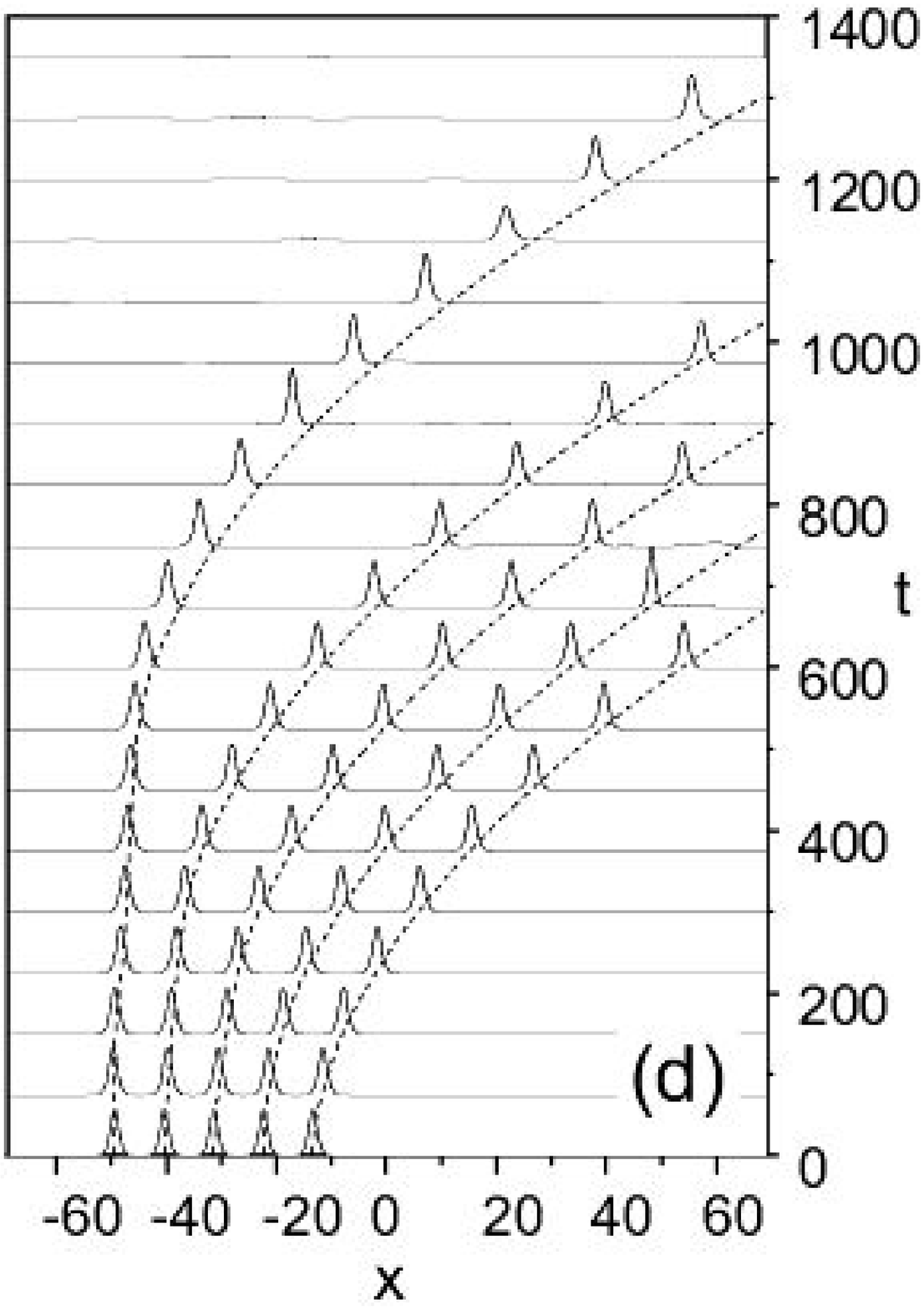}}
\caption{Controlled withdrawal of solitons from the 5-soliton
train by adjusting the strength of the linear potential
(\ref{tilted}) with parameters: $A = -0.0005$, $\Omega =2\pi/9$,
$\Omega_0 = 0$. Depending on the tilt, different number of
solitons can be pulled out of the train: (a) one soliton at $B = -
0.00003$, (b) two at $B = -0.00011$, (c) four at $B = -0.0002$ and
(d) five at $B = -0.0003$. The initial phase difference and
separation between neighboring solitons in the train are,
respectively, $\pi$ and $9$. Initially the train is shifted by
$-10 \pi$ with respect to $x=0$ for graphical convenience. Solid
and dashed lines correspond, respectively, to direct simulations
of the NLS equation (\ref{eq:nls}) and numerical integration of
the PCTC system (\ref{eq:mu-k-p}) - (\ref{eq:del-k-p}).}
\label{fig7}
\end{figure}

Controlled manipulation with matter - wave solitons is important
issue in these applications. Below we demonstrate that solitons of
attractive BEC confined in optical lattice can be flexible
manipulated by adjustment of the strength of the linear potential.
In Fig. \ref{fig7} we show the extraction of different number of
solitons from the 5-soliton train by increasing the strength of
the linear potential $B$, as obtained from direct simulations of
the NLS equation (1) and numerical integration of the PCTC system
(38) - (41) with
\begin{eqnarray}
M_k^{(2)} &=& {\pi A\Omega^2 \over 8\nu _k } {1 \over \sinh Z_k
}\sin (\Omega \xi_k +\Omega _0) - \frac{1}{2} B,\label{mk}\\
D_k^{(2)} &=&- {\pi^2 A\Omega ^2 \over 16\nu_k^2 } {\cosh Z_k
\over \sinh ^2 Z_k } \cos (\Omega \xi_k +\Omega _0)
\nonumber\\
&-&B \xi_k. \label{dk}
\end{eqnarray}

As is evident from Fig. \ref{fig7}, the PCTC model provides adequate
description of the dynamics of a N-soliton train in a tilted periodic
potential. A small divergence between predictions for the trajectory
of the left border soliton in Fig. \ref{fig7} (d), is due to the
imperfect absorption of solitons from the right end of the integration
domain. Reflected waves enter the integration domain and interact with
solitons, which causes the discrepancy.

\subsection{Periodic potential}

Another external potential in which the $N$-soliton train exhibits
interesting dynamics is the periodic potential of the form $V(x) = A
\cos(\Omega x + \Omega_0)$. This case also may have a direct relevance to
matter - wave soliton trains confined to optical lattices.
The PCTC system in terms of soliton parameters has the form:
\begin{eqnarray}\label{eq:mu-k-p}
{d\mu _k \over dt } &=& 16\nu _0^3 \left( e^{-2\nu _0 (\xi_{k+1}
-\xi_{k})} \cos \Phi _k \right. \nonumber \\ &-& \left. e^{-2\nu _0
(\xi_{k} -\xi_{k-1})} \cos \Phi _{k-1}\right) + M_k^{(2)}(\nu _k),\\
\label{eq:nu-k-p}
{d\nu _k \over dt } &=& 16\nu _0^3 \left(
e^{-2\nu _0 (\xi_{k+1} -\xi_{k})} \sin \Phi _k\right.  \nonumber\\
&& \qquad - \left. e^{-2\nu _0 (\xi_{k} -\xi_{k-1})} \sin \Phi _{k-1}
\right),\\
\label{eq:xi-k-p}
{d\xi _k \over dt } &=& 2\mu _k, \\
\label{eq:del-k-p} {d\delta _k \over dt } &=& 2(\mu _k^2 +\nu _k^2)
+ D_k^{(2)}(\nu _k),
\end{eqnarray}
where $ M_k^{(2)}(\nu _k)$, $D_k^{(2)}(\nu _k)$ are given in
(\ref{eq:P16.1}) and (\ref{eq:P16.2}), and $\Phi _k $ -- in eq.
(\ref{eq:Phi-k}).

\begin{figure}
\includegraphics[width=7cm,height=5cm,clip]{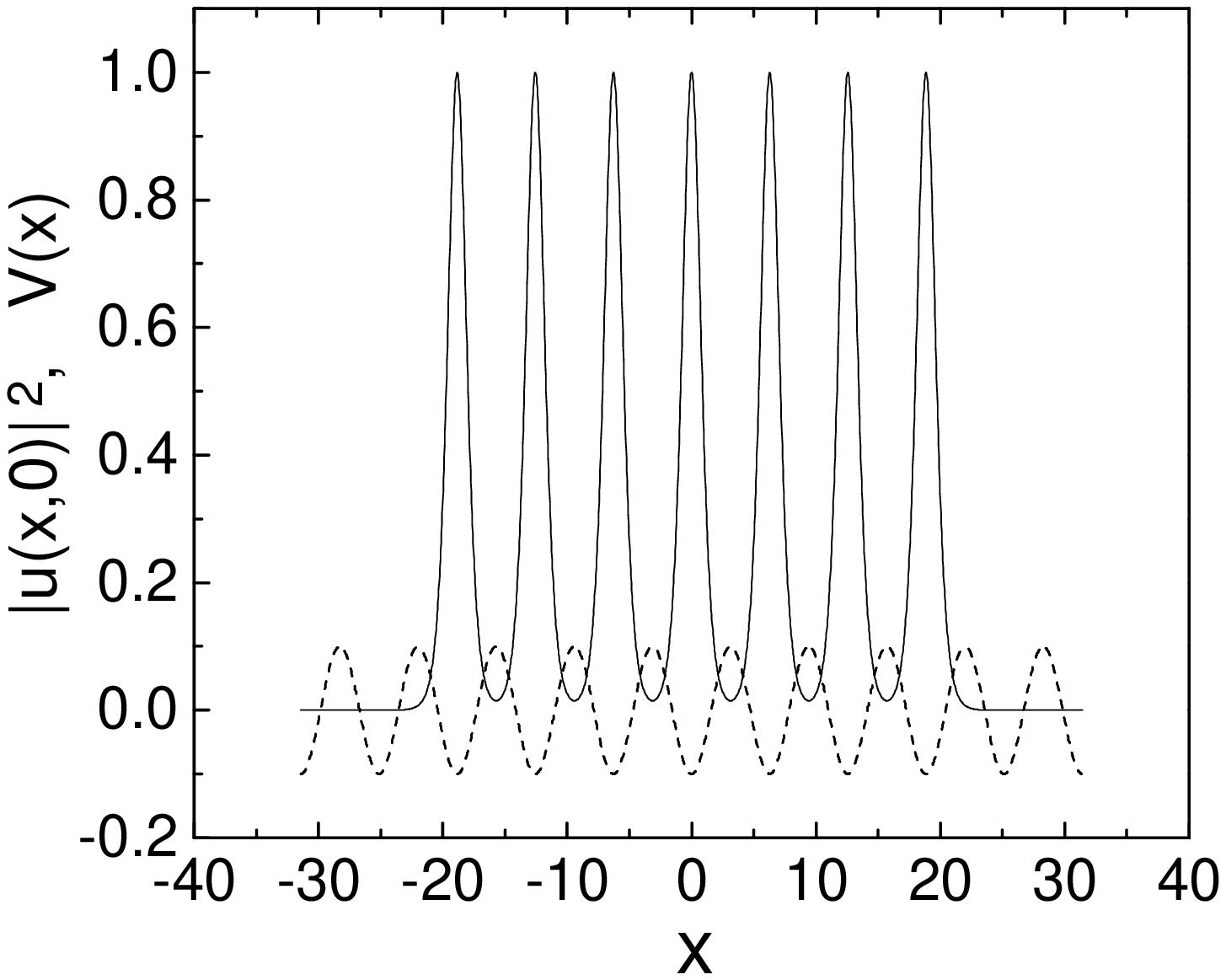}
\caption{Solitons (continuous line) remain confined around the
minima of the periodic potential $V(x) = A \cos(x)$ (dashed line)
performing small amplitude oscillations if its strength is big
enough $A=-0.1$.} \label{fig4}
\end{figure}

\begin{figure}
\includegraphics[width=3.7cm,height=5cm,clip]{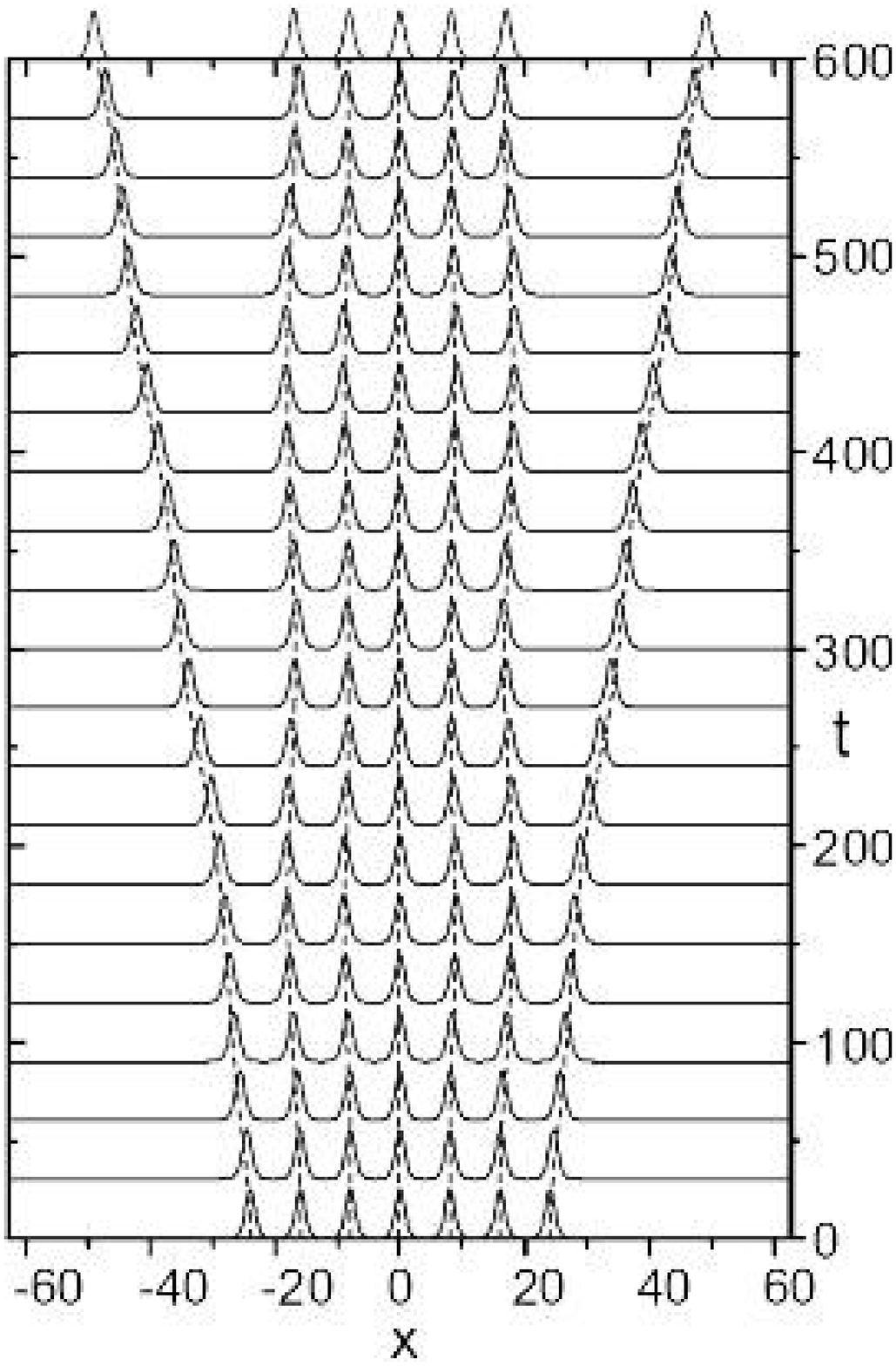} \qquad
\includegraphics[width=3.7cm,height=5cm,clip]{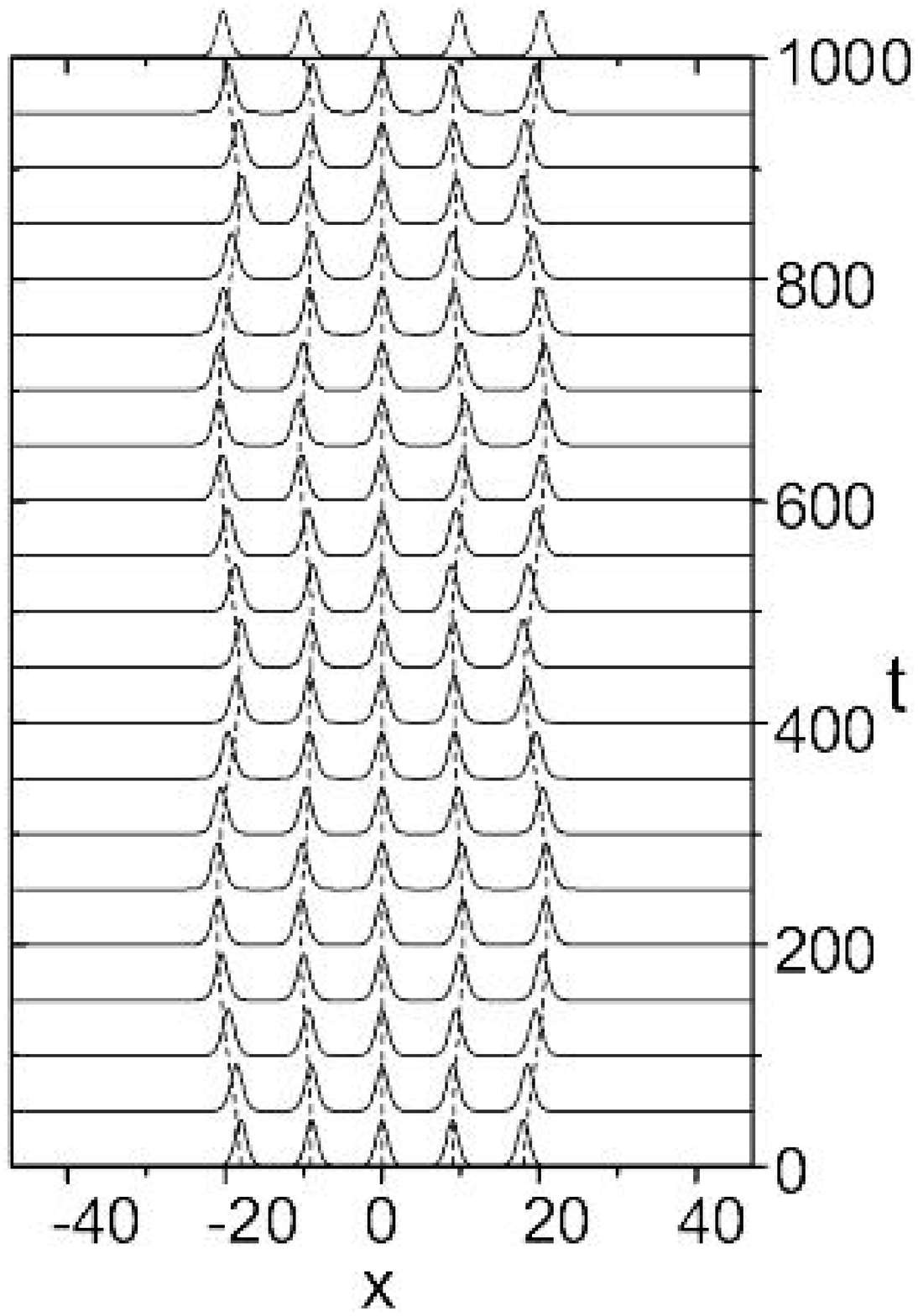}
\caption{Left panel: The expulsion of solitons from the train, as
obtained from direct simulations of the NLS equation
(\ref{eq:nls}) (solid lines), and as predicted by PCTC system
(\ref{eq:mu-k-p})-(\ref{eq:del-k-p}) for the center of mass
$\xi_i$ (dashed lines).  The IC of the $7 $-soliton train are
given by (\ref{eq:in-par}) with $r_0=8 $; the parameters of the
periodic potential $V(x) = A \cos(\Omega x + \Omega_0)$ are $A=-
0.001, \ \Omega = \pi/4, \ \Omega_0 = 0$. Right panel:
Oscillations of the 5-soliton train with IC given by
(\ref{eq:in-par}) with $r_0=9 $; in a moderately weak periodic
potential, $A = -0.0005$, $\Omega = 2\pi/9$, $\Omega _0 = 0$.
Solid and dashed lines correspond, respectively, to numerical
solution of the NLS eq. (\ref{eq:nls}) and PCTC system
(\ref{eq:mu-k-p}) - (\ref{eq:del-k-p}). } \label{fig5}
\end{figure}

Each soliton of the train experience confining force of the periodic
potential and repulsive force of neighboring solitons. Therefore,
equilibrium positions of solitons do not coincide with the minima of the
periodic potential. Solitons placed initially at minima of the periodic
potential (Fig. \ref{fig4}) perform small amplitude oscillations around
these minima, provided that the strength of the potential is big enough to
keep solitons confined. As opposed, the weak periodic potential is unable
to confine solitons, and repulsive forces between neighboring solitons (at
phase difference $\pi$) induces unbounded expansion of the train.
In the intermediate region, when the confining force of the
periodic potential is comparable with the repulsive forces of
neighboring solitons, interesting dynamics can be observed such as
the expulsion of bordering solitons from the train, as shown in
the left panel of Fig. \ref{fig5}.
This phenomenon, revealing the complexity of the internal dynamics
of the train,  can be explained as follows. Each soliton performs
nonlinear oscillations within individual potential wells under
repulsive forces from neighboring solitons. When the amplitude of
oscillations of particular solitons grow and two solitons closely
approach each other, a strong recoil momentum can cause the
soliton to leave the train, overcoming barriers of the periodic
potential. In Fig. \ref{fig5} this happens with bordering solitons
(the other  solitons remain bounded under long time evolution). It
is noteworthy to stress that this phenomenon is well described by
the PCTC model, as is evident from Fig. \ref{fig5}, left panel.

On the right panel of the same figure we have similar IC
as in (\ref{eq:in-par}) and we have choosen again
the initial positions of the solitons to coincide with the minima
of the periodic potential $V(x) = A \cos(\Omega x +\Omega_0)$; i.e.
$r_0=2\pi/\Omega  $. The values of $A=-0.0005 $ and $r_0=9 $ in the right
panel of Fig. \ref{fig5} now are such that the solitons form a bound
state. Therefore for any given initial distance $r_0 $ there is a critical
value  $A_{\rm cr}(r_0) $ for $A $ such that for $A>A_{\rm cr}(r_0) $ the
soliton train with IC (\ref{eq:in-par}) will form a bound state.

In contrast to the quadratic potentials, the weak periodic potential is
unable to confine solitons, and repulsive forces between neighboring
solitons (at $\nu _k(0) =1/2, \ \delta _k(0) =k\pi$) induces unbounded
expansion of the train similar to what was shown in the left panel of
Fig. \ref{fig1}.

\begin{figure}
\includegraphics[width=3.7cm,height=5cm,clip]{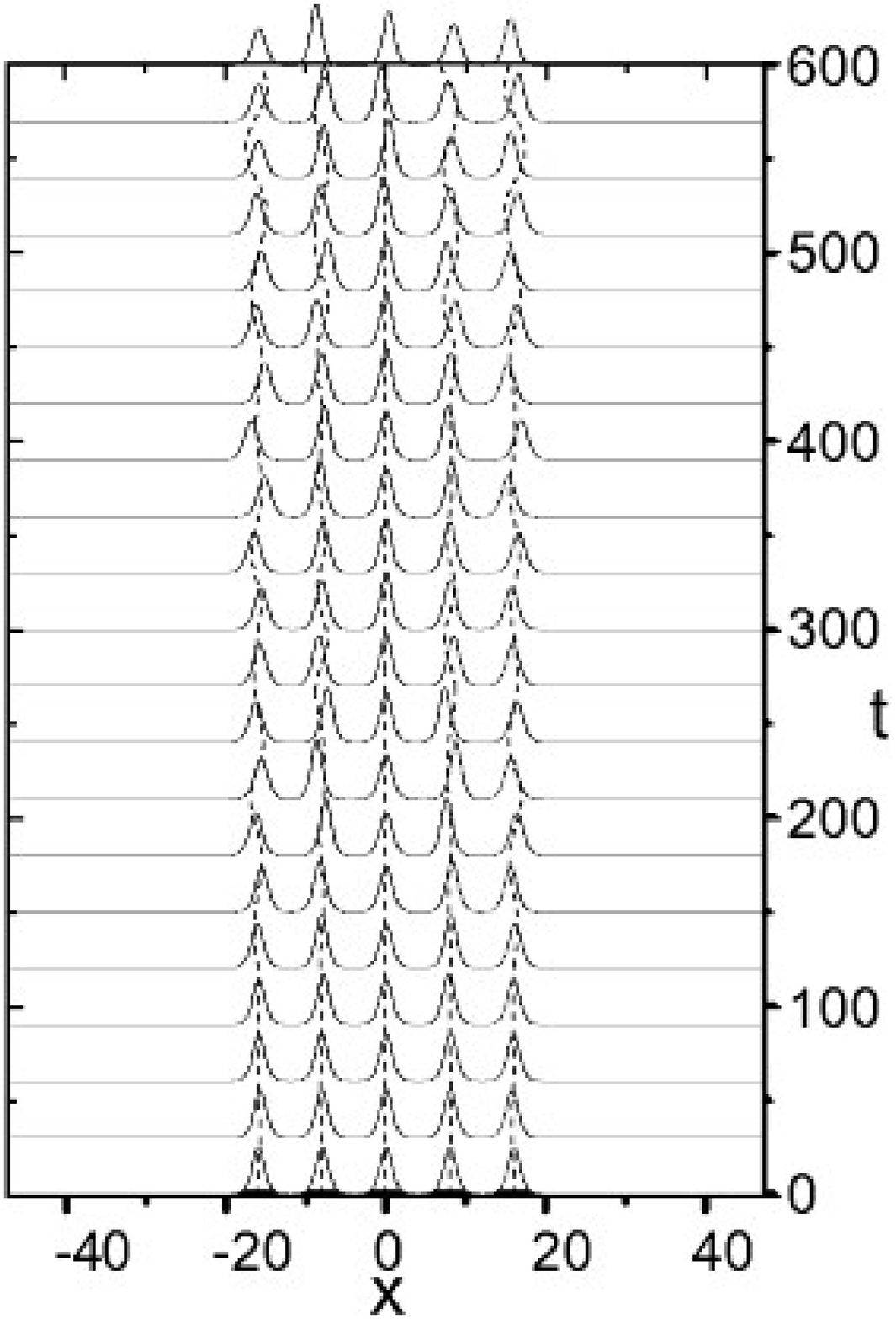}
\qquad
\includegraphics[width=3.7cm,height=5cm,clip]{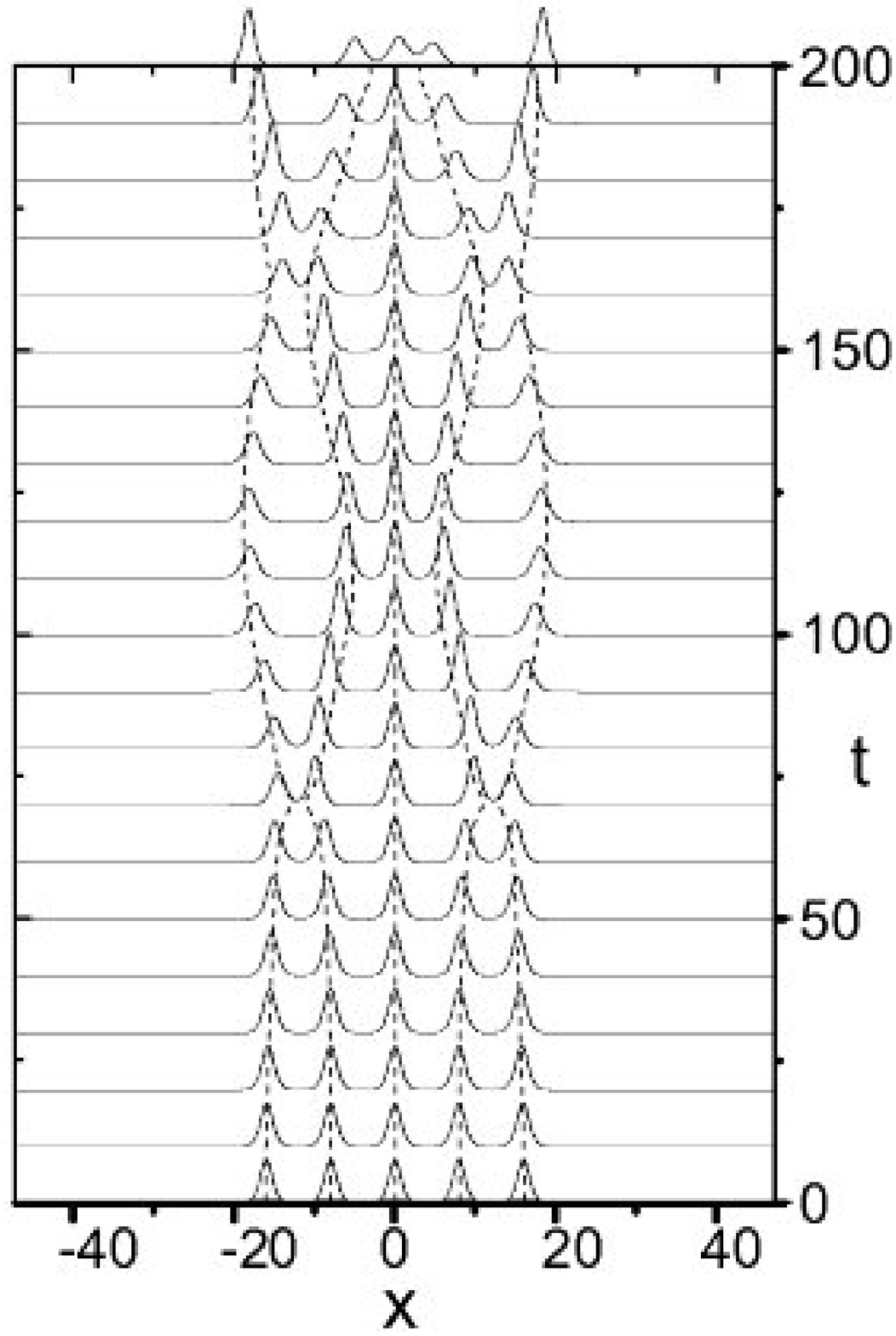}
\caption{Dynamics of a 5-soliton train with zero phase difference
between neighboring solitons, in the periodic potential $V(x) = A
\cos(\Omega x +\Omega_0)$ with $\Omega = \pi/4$, $\Omega_0=0$, and
$r_0 = 8$. Left panel: When the periodic potential is strong
enough $A = -0.02$, the N-soliton train remain confined, each
soliton performing small amplitude oscillations around the minima
of individual cells. Right panel: Weaker periodic potential
$A=-0.01$ cannot prevent solitons from collisions, which destroy
the train. Solid and dashed lines correspond, respectively, to
numerical solution of the NLS eq. (\ref{eq:nls}) and PCTC system
(\ref{eq:mu-k-p}) - (\ref{eq:del-k-p}). }
\label{fig6}
\end{figure}

The periodic potential can play stabilizing role also for the IC
(\ref{eq:in-para}), when the zero phase difference between
neighboring solitons correspond to their mutual attraction.
If the periodic potential is strong enough, solitons do not experience
collision. The weak periodic potential cannot prevent solitons from
collisions, which eventually leads to destruction of the soliton train, as
illustrated in Fig. \ref{fig6}.
Again for any given initial distance $r_0 $ there will be a critical
value $A_{\rm cr}'(r_0) $ for $A $ such that for $A>A'_{\rm cr}(r_0) $ the
soliton train with IC (\ref{eq:in-para}) will form a bound state avoiding
collisions.

\begin{figure}
\includegraphics[width=3.7cm,height=5cm,clip]{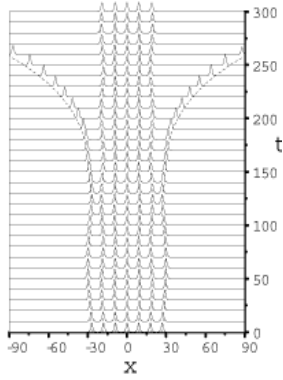}
\caption{Stabilization of a soliton train in a combined potential
(periodic $+$ inverted parabola):
$V(x) = -0.02 \cos(\Omega x) - 0.000145 x^2$. Separation between
in-phase ($\delta_k=0$) solitons is $r_0=9$, $\Omega = 2\pi/9$. Bordering
solitons leave the train as they experience stronger expulsion, while
the central ones remain bounded.}
\label{fig6c}
\end{figure}

Attractive interactions at zero phase difference between
neighboring solitons can be balanced by expulsive force on
solitons, if the train is positioned on an inverted parabolic
trap. In this case solitons far from the center experience
stronger expulsive force and leave the train, as illustrated in
Fig. \ref{fig6c}.

\section{Hamiltonian approach to perturbed NLS and CTC}\label{sec:Ham}

The Hamiltonian method has played important role in the analysis of
integrable and close to integrable nonlinear evolution equations and
dynamical systems, see \cite{FaTa}. In this section we will outline how
this method can be used for the analysis of the $N $-soliton interactions.

The relation between the Hamiltonian properties of the NLS eq. and the CTC
model was derived in \cite{VS-EPJ29}. Here we will show that this approach
can be extended also to the perturbed versions of NLS and CTC. Indeed, the
Hamiltonian of eq. (\ref{eq:nls}) with $iR[u]=V(x)u(x,t) $ is equal to:
\begin{eqnarray}\label{eq:H-nls}
H &=& H_{\rm NLS} + H_{\rm V}, \\
\label{eq:H-nls1}
H_{\rm NLS} &=& \int_{-\infty }^{\infty } dx\, {1 \over 2} \left( |u_x|^2
- |u(x,t)|^4\right) , \\
\label{eq:H-nlsV}
H_{\rm V} &=& \int_{-\infty }^{\infty } dx\, V(x) |u(x,t)|^2 ,
\end{eqnarray}

One of the ways  to derive the CTC from the NLS is based on the use of the
variational method \cite{AndLis,Arnold}.  Namely one constructs the
Lagrangian of NLS, then takes an anzatz of the form:
\begin{equation}\label{eq:uN-s}
u(x,t) = \sum_{k=1}^{N} u_{k}^{(\rm 1s)} (x,t),
\end{equation}
and integrate over $x $ neglecting terms of order $\epsilon ^{k} $ with
$k>1 $. Here $u_{k}^{(\rm 1s)} (x,t) $ is the one-soliton pulse with
parameters $\mu _k $, $\nu _k $, $\xi_k $ and $\delta _k $. The results
must depend only on the $4N $ soliton parameters. It was pointed out in
\cite{VS-EPJ29} that if we apply this method directly to the Hamiltonian
$H_{\rm NLS} $ we get additional singular in $\epsilon $ elements which
are taken care of by a proper regularization. The regularized Hamiltonian
is
\begin{eqnarray}\label{eq:H-reg}
H_{\rm reg} &=& 4(\mu _0^2+\nu _0^2) C_1 -4\mu _0C_2 +H_{\rm NLS}, \\
C_1 &=& \int_{-\infty }^{\infty } dx\,  |u(x,t)|^2, \nonumber\\
C_2 &=& \int_{-\infty }^{\infty } dx\, {i \over 2} (u^*_xu(x,t)
-u^*(x,t)u_x), \nonumber
\end{eqnarray}
Then one can show that
\begin{equation}\label{eq:H-nls-ctc}
H_{\rm reg} = 32\nu _0H_{\rm CTC} + \const
\end{equation}
where $H_{\rm CTC} $ is the Hamiltonian of the CTC. It is obtained from
the Toda chain Hamiltonian:
\begin{equation}\label{eq:H-TC}
H_{\rm TC} =  \sum_{k=1}^{N} {1 \over 2}p_k^2 + \sum_{k=1}^{N-1}
e^{q_{k+1}-q_k},
\end{equation}
by a complexification procedure after which the dynamical variables
become complex-valued:
\begin{equation}\label{eq:compl}
p_k \to P_k = p_{0,k}+ip_{1,k} , \qquad q_k \to Q_k = q_{0,k}+iq_{1,k} ,
\end{equation}
which must satisfy the Poisson brackets:
\begin{eqnarray}\label{eq:PB1}
\{p_{0,k},q_{0,s}\} =\delta _{ks}, \qquad
\{p_{1,k},q_{1,s}\} =-\delta _{ks}, \\
\{p_{0,k},q_{1,s}\} =0, \qquad \{p_{1,k},q_{0,s}\} =0,
\end{eqnarray}
Then the CTC can be written down as a standard Hamiltonian system with
$2N $-degrees of freedom and Hamiltonian provided by the real part of the
complexified $H_{\rm TC} $:
\begin{eqnarray}\label{eq:H-CTC}
H_{\rm CTC} &=& \sum_{k=1}^{N} {1 \over 2}(p_{0,k}^2 - p_{1,k}^2)
\nonumber\\
&+& 16\nu_{0}^2\sum_{k=1}^{N-1} e^{q_{0,k+1}-q_{0,k}} \cos
(q_{1,k+1}-q_{1,k}),
\end{eqnarray}
\begin{equation}\label{eq:H-CTC1}
p_{0,k} = {dq_{0,k}  \over dt } = -4\nu _0\mu _k , \qquad
p_{1,k} = {dq_{1,k}  \over dt } = -4\nu _0\nu _k.
\end{equation}
It remains to replace $q_{0,k} $ and $q_{1,k} $ in terms of the soliton
parameters in order to get the final expression for $H_{\rm CTC} $.

Let us now derive the Hamiltonian for the perturbed CTC. To this end we
will evaluate $H_{\rm V} $ in terms of the soliton parameters. Inserting
the anzats (\ref{eq:uN-s}) into the integrand for $H_{\rm V} $ we obtain
two types of terms. The first type is
\begin{eqnarray}\label{eq:Hv-1}
H_{\rm V} &=& \sum_{k=1}^{N} H_k + \sum_{k=1}^{N} (H_{k,k-1}+H_{k,k+1}),
\\
\label{eq:Hv-k}
H_k &=& \int_{-\infty }^{\infty } dx\, V(x)|u_k^{(\rm 1s)}(x,t)|^2, \\
H_{k,k-1} &=& \int_{-\infty }^{\infty } dx\, V(x) (u_k^{(\rm 1s),*}
u_{k-1}^{(\rm 1s)})(x,t). \nonumber
\end{eqnarray}

In what follows we will neglect the terms $H_{k,k-1} $ as compared to
$H_{k} $, because their ratio is of the order of $\epsilon  $.
The integrals in (\ref{eq:Hv-k}) for the quadratic and periodic potentials
have the form:
\begin{eqnarray}\label{eq:Hv-k2}
H_k &=& 4\nu _k \left( \left(V_2(\xi_k^2 - {\pi^2  \over 48\nu _k^2 }
\right) + V_1 \xi_k +V_0  \right) \nonumber\\
&+& {\pi A\Omega  \over \sinh Z_k } \cos (\Omega \xi_k +\Omega _0).
\end{eqnarray}

One can also evaluate the Poisson brackets between the soliton parameters
inserting the expressions for $p_{\alpha ,k} $ and $q_{\beta ,s} $ with
$\alpha , \beta =0,1 $ into (\ref{eq:PB1}). We skip the lengthy
calculations here and only note that these Poisson brackets combined with
the Hamiltonian
\begin{eqnarray}\label{eq:H-pert}
H_{\rm PCTC} &=& H_{\rm CTC} + \sum_{k=1}^{N} H_k\\
 &=& H_{\rm CM} + H_{\rm V},
\end{eqnarray}
indeed produce the equations of motion for the PCTC.

Our next step is to separate the center of mass motion described by
$H_{\rm CM} $:
\begin{eqnarray}\label{eq:H-cm}
H_{\rm CM} &=& N \left(8\nu _0^2 (\mu _0^2-\nu _0^2) +4\nu _0
(V_2\xi_0^2 +V_1\xi_0 +V_0)  \right. \nonumber\\
&-& \left. {\pi A\Omega \over \sinh ({\pi\Omega \over 4\nu _0}) }\cos
(\Omega \xi_0+\Omega _0)\right),
\end{eqnarray}
where the subscript 0 stands for the average value of the corresponding
parameter, e.g. $\xi_0 = 1/N \sum_{k=1}^{N}\xi_k $. Then the Hamiltonian
$H_{\rm V} $ would describe the relative motion of the solitons. We will
express it as a function of the averaged parameters $\mu _0 $, \dots,
$\delta _0 $ and the relative parameters:
\begin{eqnarray}\label{eq:t-mu}
\widetilde{\mu }_k = \mu _k -\mu _0, \qquad \widetilde{\nu }_k = \nu
_k-\nu _0, \nonumber\\
\widetilde{\xi}_k=\xi_k-\xi_0, \qquad
\widetilde{\delta }=\delta _k-\delta _0,
\end{eqnarray}
Note, that only $N-1 $ of the relative parameters are independent;
obviously they satisfy $\sum_{k=1}^{N} \widetilde{X}_k=0 $ where
$\widetilde{X}_k $ stands for $\widetilde{\mu }_k $, \dots,
$\widetilde{\delta }_k $.

In evaluating $H_{\rm V} $ we will neglect higher order terms, i.e. terms
of the order $\epsilon ^{3/2} $ and higher, as well as terms of the order
$V_s\epsilon ^{1/2} $, $A\epsilon ^{1/2} $ and higher. As a result
$H_{\rm CM} $  and $H_{\rm V} $ simplify to:
\begin{eqnarray}\label{eq:H-V}
H_{\rm CM} &=& 8\nu _0^2(\mu _0^2-\nu _0^2) +4\nu _0 \left(V_2\xi_0^2
+V_1\xi_0 +V_0 -  {V_2\pi^2  \over 48\nu _0^2 } \right) \nonumber\\
&+& {\pi A\Omega \over \sinh Z_0 } \cos (\Omega \xi_0 +\Omega _0), \\
H_{\rm V}&=& 8\nu _0^2 \sum_{k=1}^{N} (\widetilde{\mu }_k^2 -
\widetilde{\nu }_k^2 +\widetilde{H}_k ) \nonumber \\ &+&
16\nu_{0}^2\sum_{k=1}^{N-1} e^{q_{0,k+1}-q_{0,k}} \cos
(q_{1,k+1}-q_{1,k}), \\
\widetilde{H}_k &=& 4\nu _0V_2 (\xi_k -\xi_0)^2 \nonumber \\ &+&
{\pi A\Omega \over \sinh Z_0} ( \cos (\Omega \xi_k+\Omega _0) -
\cos (\Omega \xi_0+\Omega _0) ).
\end{eqnarray}

 The Hamiltonian $H_{\rm CM} $ which describes  the center of mass motion
is more simple than $H_{\rm V} $ and often one is able to solve explicitly
the corresponding equations of motion, see Section IV. The next step would
be, using the known expressions for the averaged variables $\mu _0(t) $,
\dots, $\delta _0(t) $ to insert them into $H_{\rm V} $ and try to analyze
the corresponding equations of motion. This would provide us with
information about the relative motion of the solitons around the center of
mass.  Usually we get a set of nonlinear and non-integrable ODE.

Our idea here is to use the explicit form of $H_{\rm V} $ for the
estimation of the critical values of potential strengths $A $,
$V_2 $, $V_1 $ for which the soliton motion becomes qualitatively
different. Doing this we are making use of the following
hypothesis based on the well known fact that bound states have
negative energies while asymptotically free motions should
correspond to positive energies. Therefore we evaluate the
Hamiltonian $H_{\rm V} $ inserting in it the initial soliton
parameters along with the known expressions for $\mu _0 $, \dots,
$\delta _0 $. If the result is negative we may expect that the
relative motion of the solitons will be bounded; otherwise one may
expect that at least one (or more) of the solitons will move away
from the others. The critical value of the corresponding constants
will be derived below with the condition $H_{\rm V} =0$.

Assume we have only periodic potential present and the initial soliton
configuration is (\ref{eq:in-par}). For $A=0 $  the solitons will go into
asymptotically free regime. Switching on the self-consistent periodic
potentials (such that the solitons initially are located at its minima) it
is natural to expect that for $A>A_{\rm cr} $ the solitons will be
stabilized into a bound state. Then from the condition $H_{\rm V} =0$ we
get:
\begin{equation}\label{eq:A1cr}
A_{\rm cr}= -\left(1-{1  \over N}\right) {64\nu _0^4 \over
\pi\Omega } e^{-2\nu _0r_0} \sinh {\pi\Omega \over 4\nu _0 }.
\end{equation}

Note that the critical values generically should depend not only on the
number of solitons $N $, but also on the initial configuration. The
approach we used is not very sensitive to this. It can not provide us with
the intermediary critical values when the soliton train is stabilized
after emmiting two or more solitons.

We compared the theoretical predictions for $A_{\rm cr} $ from eq.
(\ref{eq:A1cr}) with the data coming from the numeric solutions of
the corresponding PCTC for different choices of the initial
distance between the solitons $r_0 $. The results are collected
into the table \ref{tab:1}. The conclusion is that eq.
(\ref{eq:A1cr}) provides correct dependence of $A_{\rm cr} $ on
$r_0 $ up to an overall constant factor of the order of $ 1.5 $.

\begin{table}{}
\begin{tabular}{llll}
$r_0 $ & $A_{\rm cr}^{\rm exp} $ & $A_{\rm cr}^{\rm th} $ &
$A_{\rm cr}^{\rm exp} /A_{\rm cr}^{\rm th} $ \\ \hline
$2\pi $ & -0.0053 & -0.00365 & 1.45 \\
7 & -0.0025 & -0.00166 & 1.51 \\
8 & -0.00084 & -0.00057 & 1.47 \\
9 & -0.00030 & -0.00020 & 1.50
\end{tabular}
\caption{The values of $A_{\rm cr}^{\rm exp} $ obtained from
numeric simulations with PCTC versus $A_{\rm cr}^{\rm th} $
obtained from eq. (\ref{eq:A1cr}) for $N=3 $. \label{tab:1}}
\end{table}

In the next table \ref{tab:2} we summarize two sets of critical
values for the $5$-soliton trains. From our numeric experiment we
derive two critical values of $A$. The first one $A_{\rm cr}^{\rm
5,exp} $ describes the value of $A$ above which {\em all $5$
solitons} form a bound state; the second one $A_{\rm cr}^{\rm
3,exp} $ shows the value of $A$ above which the three middle
solitons form a bound state while the two end ones separate.

\begin{table}{}
\begin{tabular}{llllll}
$r_0 $ & $A_{\rm cr}^{\rm 5, exp} $ & $A_{\rm cr}^{\rm 3, exp} $ &
$A_{\rm cr}^{\rm th} $ & $A_{\rm 5,cr}^{\rm exp} /A_{\rm cr}^{\rm
th} $ & $A_{\rm 3,cr}^{\rm exp} /A_{\rm cr}^{\rm th} $ \\ \hline
$2\pi $ & -0.0068 & -0.0029 & -0.0044 &  1.55 & 0.66 \\
7 & -0.0031 & -0.0013 & -0.0020 & 1.56 & 0.65 \\
8 & -0.00115 & -0.00043 & -0.00068 & 1.68 & 0.63 \\
9 & -0.00041 & -0.000155  & - 0.00024 & 1.71 & 0.65
\end{tabular}
\caption{The values of $A_{\rm cr}^{\rm 5, exp} $ and $A_{\rm
cr}^{\rm 3, exp} $ obtained from numeric simulations with PCTC
versus $A_{\rm cr}^{\rm th} $ obtained from eq. (\ref{eq:A1cr})
for $N=5 $. \label{tab:2}}
\end{table}
Again we see, that formula  (\ref{eq:A1cr}) describes correctly
the dependence of both critical values on $r_0$.

\section{Conclusions} \label{sec:PCTC}

We have studied the dynamics of a $N$-soliton train confined to
external fields (quadratic, periodic and tilted potentials). Both
the analytical treatment in the framework of the PCTC model, and
numerical analysis by direct simulations of the underlying NLS
equation show that the PCTC is adequate for description of the
adiabatic $N$-soliton interactions in weak external potentials.
Hamiltonian approach for the perturbed NLS and CTC has been
developed, and applied to the analysis of the soliton "expulsion"
from the train, which is confined to a periodic potential.

As a physical system of direct relevance we have considered
matter-wave soliton trains in magnetic traps and optical lattices.
In the range of parameters for the trapping potential, used in the
BEC soliton train experiments, we found a good agreement between
the analytical estimates based on the PCTC model and numerical
simulations of the governing NLS equation. In what concerns the
critical strength of the periodic potential at which the soliton
expulsion occurs, analytical predictions qualitatively agree with
numerical simulations. The developed theory can be useful for
controlled manipulation with matter-wave soliton trains.
\begin{acknowledgments}

V.S.G. and N.A.K. acknowledge partial support from the Bulgarian
Science Foundation through contract No. F-1410. B.B. B. wishes to
thank the Department of Physics of the University of Salerno,
Italy, for providing a research grant during which part of this
work was done. M. S. acknowledges financial support from the MIUR,
through the inter-university project PRIN-2003, and from the
Istituto Nazionale di Fisica Nucleare, sezione di Salerno. We are
grateful to Prof. I. Uzunov for useful discussions and for calling
our attention to Refs. \cite{WB,UGL}.

\end{acknowledgments}

\end{document}